\documentclass[journal]{IEEEtran}
\usepackage[T1]{fontenc}

\usepackage[latin5]{inputenc}
\usepackage{times}
\usepackage[figuresright]{rotating}
\usepackage{float}
\usepackage{subcaption}
\usepackage{url}
\usepackage{setspace}
\usepackage{balance}
\usepackage{multirow}
\usepackage{color}
\usepackage{graphicx}
\usepackage[implicit=false]{hyperref}
\usepackage{epsfig}
\usepackage{amsmath, amssymb}

\usepackage{capt-of}
\usepackage{multirow}
\usepackage{array}
\usepackage{caption} 
\usepackage{pslatex}
\usepackage{enumitem}
\usepackage{mathtools}
\usepackage{listings}
\definecolor{codegreen}{rgb}{0,0.6,0}
\definecolor{codegray}{rgb}{0.5,0.5,0.5}
\definecolor{codepurple}{rgb}{0.58,0,0.82}
\definecolor{backcolour}{rgb}{0.95,0.95,0.92}

\lstdefinestyle{mystyle}{
    backgroundcolor=\color{backcolour},   
    commentstyle=\color{codegreen},
    keywordstyle=\color{magenta},
    numberstyle=\tiny\color{codegray},
    stringstyle=\color{codepurple},
    basicstyle=\ttfamily\footnotesize,
    breakatwhitespace=false,         
    breaklines=true,                 
    captionpos=b,                    
    keepspaces=true,                 
    numbers=left,                    
    numbersep=5pt,                  
    showspaces=false,                
    showstringspaces=false,
    showtabs=false,                  
    tabsize=2
}

\makeatletter
\newcommand{\thickhline}{%
    \noalign {\ifnum 0=`}\fi \hrule height 1.5pt
    \futurelet \reserved@a \@xhline
}

\usepackage[dvipsnames]{xcolor}
\usepackage{tcolorbox}
\usepackage{tabularx}
\tcbset{tab2/.style={colback=SkyBlue!5!white,colframe=blue!50!black,colbacktitle=Blue!40!white,
coltitle=black,center title}}

\usepackage{colortbl}

\newcolumntype{"}{@{\hskip\tabcolsep\vrule width 1pt\hskip\tabcolsep}}
\makeatother
\newcolumntype{?}{!{\vrule width 1.5pt}}
\ifCLASSOPTIONcompsoc
  \usepackage[nocompress]{cite}
  \else
  \usepackage{cite}
\fi
\ifCLASSINFOpdf
\else
\fi
\begin{document}

\newcommand{\squeezeupppp}{\vspace{-8 mm}}
\newcommand{\squeezeuppp}{\vspace{-6 mm}}
\newcommand{\squeezeupp}{\vspace{-5 mm}}
\newcommand{\squeezeup}{\vspace{-3 mm}}
\newcommand{\squeezeu}{\vspace{-2 mm}}
\newcommand{\squeeze}{\vspace{-1 mm}}
%


\title{Design of Fieldable Cross-Layer Optimized Network using Embedded Software Defined Radios: \\ Survey and Novel Architecture with Field Trials
}

\author{\IEEEauthorblockN{Jithin Jagannath, Anu Jagannath, Justin Henney, Tyler Gwin, Zackary Kane, Noor Biswas, Andrew Drozd\\}
\IEEEauthorblockA{Marconi-Rosenblatt AI/ML Innovation Lab,\\ ANDRO Computational Solutions, LLC, Rome NY \\ \{jjagannath, ajagannath, jhenney, tgwin, zkane, nbiswas, adrozd\}@androcs.com
}}

\maketitle


\begin{abstract}

The proliferation of wireless devices and their ever increasing influence on our day-to-day life is very evident and seems irreplaceable. This exponential growth in demand, both in terms of the number of devices and Quality of Service (QoS) had spawned the concept of cross-layer optimization several years ago. The primary goal of the cross-layer approach was to liberate the strict boundary between the layers of the traditional Open Systems Interconnection (OSI) protocol stack. The objective was to enable information flow between layers which then can be leveraged to optimize the network's performance across the layers mitigating the challenges due to scarce resources while meeting QoS demands. The initial decade focused on establishing the theoretical feasibility of this revolutionary concept and gauging the effectiveness and limits of this idea. During the next phase, the advent of software defined radios (SDR) accelerated the growth of this domain due to its added flexibility. Even with the immense interest and progress in this area of research, there has been a gaping abyss between solutions designed in theory and ones deployed in practice. To establish this, we first present an elaborate survey of the cross-layer protocol stack literature with emphasis on their maturity scale. Next, we briefly discuss how a commercial off-the-shelf (COTS), low SWaP (Size, Weight, and Power) embedded SDR (e-SDR) was transformed into a standalone, fieldable transceiver. Thereafter, we provide the software design ethos that focuses on efficiency and flexibility such that the optimization objectives and cross-layer interactions can be reconfigured rapidly. To demonstrate our claims, we provide results from extensive outdoor over-the-air experiments in various settings with up to 10-node network topologies. The results from the field trials demonstrate high reliability, throughput, and dynamic routing capability. To the best of our knowledge, this is the first time in literature, a COTS e-SDR has been leveraged to successfully design a cross-layer optimized transceiver that is capable of forming an ad hoc network that provides high throughput and high reliability in a ruggedized, weatherized, and fieldable form factor. 

\end{abstract}




%
\IEEEpeerreviewmaketitle

\section{Introduction}\label{sec:introduction}


In recent years, we have seen explosive growth in the number of wireless devices that have become an inevitable component of our daily lives. This includes everything such as the connected devices in our smart homes, cellular network, the entire concept of Internet-of-Thing (IoT) networks controlling manufacturing, monitoring smart grids, space communications, underwater networks, tactical networks, among others. As we move from 5G (5th Generation) to 6G (6th Generation), the need to optimize the scarce resources is becoming evident and inevitable \cite{6Genesis,6Gcomms,jagannath2020redefining, 6Gfrontier,6Greq,6Gwhat,6gsaad}. We have also seen a drive towards the software virtualization of these devices and networks to provide much-required flexibility and adaptability to the growing needs of each application. The inability to adapt the network operations depending on varying requirements and dynamic deployed scenarios due to black-box style wireless stack leads to inefficient resource usage and suboptimal performance. Dynamic network control and agile management (user scheduling, radio resources, mobility management) are the envisioned benefits of the next-generation wireless network paradigms such as software-defined networking (SDN), network function virtualization (NFV), and multi-access edge computing (MEC) \cite{bonati2020open}. Therefore, a reconfigurable radio stack that can be transformed based on diverse applications as well as to adapt to the dynamic wireless conditions is favored.

Traditionally, the strictly layered architecture proposed by the open systems interconnection (OSI) reference model has been the prevalent design for a majority if not all modern networking architectures. This is strict in the sense that they are designed to maintain only a limited interface between the neighboring layers \cite{Fu_survey}. Realizing the deficiencies in this layered architecture, the cross-layer optimized approach has been proposed over the past decade to maximize the utilization of scarce resources by "erasing" the strict boundaries between various layers of the protocol stack. In other words, any attempt to violate the OSI reference model is considered a cross-layer design \cite{Foukalas_survey}. While there are abundant solutions proposed in literature \cite{Fu_survey,cross_survey_1,cross_survey_2,cross_survey_5_tutorial, Rappaport_cross_survey}, the majority of it is limited to simulations that may have strong assumptions and/or do not consider all the hardware constraints and rigidness that may be encountered during a real-life deployment. During the next phase of advancement, the advent of software defined radios (SDR) provided the much-needed impetus to this concept providing the flexibility to implement novel cross-layer architectures. This enabled some of these efforts to be extended to hardware-based testbed evaluations. In most cases, these efforts still used one or more dedicated (non-embedded) host computers to implement the solutions which were then connected to SDRs. 

\subsection{Motivation, Problem Statement, and Contribution}

\textbf{Motivation:} Even with the advances discussed above, to the best of our knowledge, there does not exist a ruggedized and fieldable SDR with a comprehensive cross-layer optimization capable software module implemented on an embedded ARM processor. The main reason for this is the various hurdles that are associated with developing the solution from theory to effective hardware deployable software. In this article, we aim to demonstrate that this challenge can be overcome using a Commercial off-the-shelf (COTS) embedded SDR (e-SDR) and a novel efficient software architecture. Therefore, in this article, we present the first, completely stand-alone, ruggedized, and fieldable cross-layer optimized solution built using a low SWaP (Size, Weight, and Power) e-SDR.

\textbf{Problem Statement:} Driven by our motivation, we define a specific problem statement that is considered in this work. Design a fieldable e-SDR with an energy-aware cross-layer optimized protocol that aims to maximize network lifetime for enabling telemetry collection of tactical test and evaluation ranges.

The primary objective of this work is the transition of theoretical cross-layer concepts that have been shown to be effective in simulations to a fieldable hardware entity. Our implementation principle has been to facilitate rapid reconfiguration of network objectives with only a few lines of code thereby enabling a \emph{truly} software-defined radio that evolves with the growing requirements. To accomplish this, in the extension to our prior work \cite{JagannathMILCOM2021}, we first discuss the hardware-level modifications required to customize the baseline e-SDR into a fieldable solution. Next, we outline how the software protocol stack was designed and implemented on a computationally constrained ARM processor. To demonstrate the feasibility of the implementation and the designed transceiver in terms of throughput and reliability, we performed extensive outdoor experiments with up to 10 nodes in the network. The proposed architecture and design principles can be leveraged to implement and mature several of the novel cross-layer optimized solutions to meet the evolving needs of both tactical and commercial communication systems. We hope and believe the unique cross-layer design methodologies and extensive outdoor evaluations (field trials) will serve as an impetus for the maturation of novel cross-layer solutions.

\textbf{Contribution:} The key contributions of the article are as follows; 
\begin{itemize}
    \item To place the contribution in perspective, we survey the domain of cross-layer optimization with an emphasis on the maturity of the solutions. specifically, we discuss how the majority of the works in this domain is restricted to either simulations or preliminary in-lab hardware analysis. 
    \item We provide a brief account of hardware components and their configuration that were used to customize the COTS e-SDR to ensure a standalone, portable solution that would provide extended range and is capable of running cross-layer optimized geographical ad hoc routing. 
    \item Embedded software-defined cross-layer protocol stack implementation with a clearly laid out explanation showing all steps from the design concept to software implementation.
    \item Additionally, we provide details of the implementation of graphical user interface (GUI) as an example of how the software protocol stack also provisions end-user-defined applications.
    \item We have conducted an elaborate outdoor experimental evaluation that includes, peer-to-peer setting, line network, 5-node topology, and 10-node topology.
    \item The experiments demonstrated long-range, high throughput, and reliability in all the settings. The real-time adaptability and dynamic routing capability of the cross-layer optimized network have been demonstrated. Finally, we also demonstrated how the network can handle multiple sessions without degradation.
\end{itemize}

The rest of the paper is organized as follows, in Section \ref{Sec:Survey}, we survey the recent works in the cross-layer domain with an emphasis on the maturity of the proposed solutions. Next, in Section \ref{Sec:SystemDesign}, we discuss the system design from the hardware and software point of view. We elaborate on the network topologies and various test objectives for the field trials that showcase the reliability, throughput, and dynamic routing capabilities of the fieldable e-SDR in Section \ref{Sec:Resutls}. Finally, we conclude the article in Section \ref{Sec:Conclusion}.

\section{Survey of Cross-layer Approaches}\label{Sec:Survey}

Cross-layer approaches have been explored as a novel solution to address a wide range of problems in wireless communication due to its perceived benefits in sustaining communications in a dynamic and constrained environment \cite{Fu_survey, Foukalas_survey, cross_survey_1, cross_survey_2, jagannath2019phd, cross_survey_3,  cross_survey_5_tutorial, Rappaport_cross_survey}. This includes tactical network \cite{Wang_cross, Nosheen_cross_OMNET, DRS}, commercial terrestrial network \cite{Zhu_TCP_cross, Herrmann_energy_cross, Jagannath19CCNC_1, Barmpounakis_crosslayer_5G}, space networks \cite{zhang_cross_space, AJagannath19CCAA}, acoustic underwater networks \cite{Zhou_cross, pompili_cross_UW, kuo_cross_UW, liu_cross_UW, vuran_cross_UW, jornet_cross_UW}, and even in the upcoming visible light communication networks \cite{Nancy18ADH, Demir_VLC, Jagannath19WOWMoM, dang_cross_visible, Jagannath18ICNC}. These solutions are designed for various objectives such as optimizing throughput and/or latency \cite{ROSA, ROCH, She_URLLC, Jagannath18TMC}, fairness \cite{Wang_cross, Park_cross_fairness, Xu_cross_fairness}, energy consumption \cite{Dai_energy_cross, Salameh_cross_energy, Al-Jemeli_energy_cross, Herrmann_energy_cross, Nosheen_cross_OMNET}, resource management \cite{Barmpounakis_crosslayer_5G, Demir_VLC}, efficient multipath TCP \cite{Zhu_TCP_cross}. The goal of this survey is not to cover the entirety of cross-layer approaches, but rather to demonstrate the gamut of the areas of application for cross-layer approaches to emphasize the impact and importance of these solutions. In contrast to other surveys, in this section, we specifically focus on reviewing the maturity of these state-of-the-art cross-layer solutions to \textit{expose the absence of a deployable e-SDR based solution}. We divide the discussion into two broad categories, works that are limited to simulations and others that have provided some form of preliminary hardware testbed-based evaluation to validate their work.  

\subsection{Initial Proof of Concept in Simulation}

As in any novel research, simulations are the obvious first choice to establish feasibility and performance gains over existing approaches. To this end, the majority of the works that propose cross-layer optimization have been limited to simulations \cite{ROSA,DRS,Herrmann_energy_cross, Wang_cross,Nosheen_cross_OMNET,Zhu_TCP_cross,ROCH, She_URLLC} due to the challenges, time, and effort it takes to evaluate them on hardware testbed. Most works in the cross-layer domain rely on MATLAB \cite{ROSA, Ekici_delay-guaranteed, DRS,Salameh_cross_energy}, NS2/3 \cite{Park_cross_fairness, Xu_cross_fairness, Al-Jemeli_energy_cross, Herrmann_energy_cross}, OMNET \cite{Nosheen_cross_OMNET} or similar wireless network simulators. In most cases, these simulations are executed under various assumptions and/or are abstracted from the physical layer (PHY) of the protocol stack and are restricted to packet-level simulators. These assumptions and abstraction imply several intricacies of real-world deployment (operating channel conditions, computational resources, RF frontend capabilities, flexibility, and associated latency/overhead) are overlooked or set aside to be handled in the future which often thwarts the maturation level of cross-layer solutions. 

Since there are a large number of solutions proposed in this domain and evaluated in simulations, we categorize them based on the objective of the cross-layer approach. At the same time, we would like to point out that these are not strict classifications as many cross-layer approaches take into account multiple metrics either in their objective function or constraints. Hence, a single work could arguably fall into multiple categories simultaneously. Therefore, in most cases, the categorization is based on at least one of its core objectives. The main purpose of the categorization is to make it easier for readers to assimilate all the content and show the widespread use of these techniques. 

\subsubsection{Maximizing Throughput}

From the very early days, throughput has been a central metric for any kind of communication device. The evolution of cellular networks, WiFi, or any other radio access technology is often driven by the desire to drastically improve the throughput of the network. Hence, it is not surprising to see a number of cross-layer optimized solutions have been developed for this exact purpose. 

Cognitive radio networks (CRNs) are referred to as networks that are capable of sensing and dynamically accessing the spectrum of interest when it is idle and not being used by primary users. Primary users are entitled to a license to use the spectrum at the highest priority. The CRN can be seen as secondary users who are allowed to access idle parts of the spectrum without disrupting the primary user's activities. These are general definitions and could be more stringently defined from case to case in actual deployments. A distributed cross-layer approach to maximize the network throughput of CRN was proposed in \cite{ROSA}. The goal was to perform opportunistic spectrum access and dynamic routing algorithm and was referred to as ROSA (ROuting and Spectrum Allocation algorithm). The algorithm enabled individual nodes to perform joint routing, dynamic spectrum allocation, scheduling, and transmit power control. The proposed solution is shown to outperform approaches that considered only spectrum allocation or routing individually using a packet-level MATLAB simulator. 

In the case of a CRN employing Code Division Multiple Access (CDMA) at the physical layer, a cross-layer algorithm to maximize throughput was proposed using a new spread-spectrum management paradigm using \cite{ROCH}. The algorithm referred to as ROCH (Routing and cOde division CHannelization) proposed jointly-designed routing and code division channelization similar to ROSA with the added dimension of codeword optimization. This work also relied on packet-level simulators for the initial proof-of-concept. 

In \cite{Liu_energy_throughput_cross}, the authors propose a cross-layer approach that utilizes parameters from Medium Access Control (MAC) and PHY layer of the IEEE 802.11p protocol in  Dedicated Short Range Communication (DSRC) based vehicular communication networks. To this end, they introduce cross-layer optimization modeling for real-time, on-road multimedia services and present a throughput-maximal framework that is designed for varying fading channel in mobile nodes with high velocities. The authors claim the performance in the simulation provided evidence for performance improvement which could be relevant even beyond vehicular networks. 

In \cite{JEMILI_multipath}, the authors propose a cross-layer multi-path routing approach that utilizes non-correlated and node-disjoint paths able to concurrently transport multimedia content from sources to the destination by accounting for contextual information. To accomplish this the nodes utilize cross-layer interaction between the MAC and the network layer to tune their wake-up schedules and determine the disjoint paths. The work is limited to simulation but the authors have observed performance improvement in terms of throughput, latency ad energy usage. 

\subsubsection{Minimizing Delay or Latency}

A key factor that impacts cellular networks and wireless mesh networks is the delay or latency experienced by the network. Latency is often a key Quality of Service (QoS) metric required by end customers relying on wireless services for the operations. Accordingly, cross-layer approaches are proposed to minimize end-to-end delay for a multihop wireless network in \cite{Cheng_cross_Delay}. The authors propose two cross-layer schemes that they refer to as, \textit{loosely coupled cross-layer scheme} and a \textit{tightly coupled cross-layer scheme}. In the loosely coupled cross-layer scheme, routing is computed first and then the information of routing is used for link layer scheduling whereas, in the tightly coupled scheme, routing and link scheduling are jointly solved in one optimization model. They showcased the superiority of their cross-layer approaches over traditional non-cross-layer approaches through simulations.

One of the key objectives of 5G networks is to support ultra-reliable and low-latency communications (URLLC). Cross-layer approaches have been leveraged recently to enable URLLC in radio access networks by devising a resource allocation policy and a proactive packet dropping strategy \cite{She_URLLC}. In this work, the authors optimize the packet dropping policy, power allocation policy, and bandwidth allocation policy to minimize the transmit power under the Quality of Service (QoS) constraint. The feasibility of the proposed approach is validated using simulations. 

In \cite{Ekici_delay-aware}, the authors identify that the end-to-end delay performance has a complex dependence on the higher-order statistics of cross-layer algorithms, and hence optimization-based design methodologies that optimize the long
term network utilization are not ideal for delay-aware design. To overcome this, the authors design a delay-aware joint flow control, routing, and scheduling algorithm for multihop wireless networks to maximize network utilization. The proposed cross-layer approach utilizes a regulated scheduling strategy based on a token-based service discipline, for shaping the per-hop delay distribution. Through simulation, the authors show optimal network utilization and better end-to-end delay performance for the proposed joint flow control, routing, and scheduling algorithm.

A cross-layer design is proposed in \cite{Ekici_delay-guaranteed} using virtual queue structures to guarantee finite buffer size or worst-case delay performance. The algorithm solves a joint congestion control, routing, and scheduling problem in a multihop wireless network while ensuring average end-to-end delay constraints per-flow and minimum data rate requirements. Through MATLAB simulations they illustrate a tradeoff between the throughput and average end-to-end delay bound while satisfying the minimum data rate requirements for individual flows.

DRS (Distributed Deadline-Based Joint Routing and Spectrum Allocation) \cite{DRS} was introduced to mitigate the last packet problem that has been shown to be a shortcoming of queue length-based backpressure scheduling algorithms such as ROSA. In this work, the authors proposed the use of a virtual queue such that the virtual queue length was designed to mitigate the effects of the last packet problem without substantial degradation of the throughput performance. The virtual queue length considered parameters such as length of the packet, remaining lifetime, and estimated time to the destination. The distributed implementation of DRS was shown to outperform ROSA in terms of effective throughput and reliability using a packet-level simulator. The reason we have this work categorized under latency is due to the definition of effective throughput used in this work. Only packets received at the destination before the deadline contributed to effective throughput which essentially is trying to ensure packet delay constraints. The simulations in MATLAB demonstrate that DRS outperforms ROSA both in terms of effective throughput and reliability.

\subsubsection{Fairness}

Two cross-layer algorithms, a dual-based algorithm, and a penalty-based algorithm are proposed in \cite{Wang_cross} to solve the rate control problem in a multihop random access network. Both algorithms can be implemented in a distributed manner, and work at the link layer to adjust link attempt probabilities and at the transport layer to adjust session rates. Their convergence and effectiveness of the proposed solution were established using simulations. 

In \cite{Park_cross_fairness}, the authors investigate the issue of fairness between IEEE 802.11-compliant wireless local area networks (WLANs) stations using Transport Control Protocol (TCP). Specifically, focusing on send/receive TCP traffic in WiFi hot spots, they show  WiFi hot spot provides more service to the wireless sending station compared to the receiving stations. This unfair service implies that the wireless sending stations dominates the use of network bandwidth restricting the access to the receiving stations. To mitigate this unfair service issue, the authors propose a cross-layer feedback mechanism in which the MAC layer at the access points measures the per-station channel utilization and system-wide channel utilization to calculate the channel access cost. Next, the TCP senders use the cost in order to assure per-station fairness and to maximize channel utilization simultaneously hence invoking a cross-layer interaction between the data link and transport layer. Their simulations in NS2 demonstrate that the proposed approach outperforms existing schemes with respect to fairness, delay, and channel utilization.

Stream Control Transmission Protocol (SCTP)-based Concurrent Multipath Transfer (CMT) was designed to improve the wireless video delivery performance with its parallel transmission and bandwidth aggregation features. In \cite{Xu_cross_fairness}, the authors point out that the shortcoming in the existing CMT solutions deployed at the transport layer is due to uncertainties at the lower layer (for example caused by variations of the wireless channel). They also point out that CMT-based video transmission may unfairly use excessive bandwidth in comparison with the popular TCP-based flows. To overcome these issues, the authors propose a fairness-driven cross-Layer SCTP-based Concurrent Multipath Transfer approach (CMT-CL/FD). CMT-CL/FD monitors and analyzes path quality, which includes wireless channel measurements at the data-link layer and rate/bandwidth estimations at the transport layer. They also propose a window-based mechanism for flow control to obtain an acceptable tradeoff between delivery fairness and efficiency. They utilize simulations to demonstrate the effectiveness of the proposed approach over traditional solutions.

\subsubsection{Minimizing Energy Utilization} In recent years, green communication protocols have gained substantial interest. The term "Green" in most cases refers to communication protocols that aim to reduce the energy consumption of wireless networks. One can imagine the implications and benefits of minimizing energy consumption in various scenarios in the context of operating cost, mitigate the need to re-deploy nodes that rely on battery, and its implication on climate change. Due to these reasons, cross-layer optimization has been leveraged as a tool for minimizing energy consumption or maximizing the network lifetime. Typically, network lifetime is defined as the total operation duration of the network before the first node depletes its energy resources. 

Authors propose a joint cross-layer optimization scheme that considers modulation, power control, and routing to maximize the energy efficiency of a wireless network \cite{Dai_energy_cross}. The constraints considered in their optimization problem include Bit Error Rate (BER) and the requested data rate. Each node in their multihop wireless network is capable of transmitting seven different modulations schemes. They use an event-driven, link-level simulator implemented using C++ to demonstrate improvement over Ad hoc On-demand Distance Vector (AODV) routing protocol. 

As discussed previously, CRN aims to use an idle (unused) portion of the licensed spectrum to access the channel in a dynamic manner. This technology is often critical for IoT devices that may be energy-constrained. To overcome this challenge, the authors of \cite{Salameh_cross_energy} propose a cross-layer approach that jointly considered the modulation order at the physical layer and the backoff probability from the MAC-layer to minimizing the energy consumption in the CR-based green IoT networks. The constraints considered in this work are to ensure IoT delay guarantees, licensed primary radio channel availability, and PR user activities. The authors use MATLAB simulations to evaluate the performance of the proposed solution. The simulations show that the proposed cross-layer design is able to reduce energy consumption while achieving the delay requirements in the network compared to traditional approaches that operate independently at the physical layer or data link (MAC) layer.

One of the key challenges of  Wireless Sensor Network (WSN) is ensuring energy efficiency while maintaining the required network performance metrics to support delay-sensitive applications. This issue is further exacerbated in networks where the sensors follow a sleep-wake duty cycle to conserve energy. To overcome some of these challenges, \cite{JEMILI_energy} propose a cross-layer multi-path routing approach designed to support duty-cycled WSN networks. Similar to \cite{JEMILI_multipath}, using the interaction between MAC and the network layer, this approach also establishes multi-node disjoint paths with complementary duty-cycling. The authors performed analytical study that proved the correctness of the proposed solutions. Simulations were able to demonstrate up to 65\% cumulative energy saving when compared to the always-on multipath approach incurring 10\% increase in response time delays. 

In \cite{Lahane2021SecuredCC}, authors propose cross-layer clustering based routing protocol for WSN to extend network lifetime. In this article, the author select cluster head based on energy consumption, delay, throughput, security, distance and overhead. The authors rely on MATLAB simulations to show how the proposed solution outperforms traditional approaches such as ant lion optimization approach and grouped grey wolf search optimization.

In order to improve the energy efficiency of the IEEE 802.15.4 based network, the authors use a cross-layer approach to control the transmission power and minimize the broadcast of control packets \cite{Al-Jemeli_energy_cross}. The authors use information from the application layer, network layer, data-link layer, and physical layer to minimize the channel occupancy time by reducing control packets (used for neighbor discovery). The proposed approach is evaluated using NS2 to achieve approximately $10\%$ reduction in energy consumption while maintaining end-to-end delays and comparable packet delivery ratios.

Another work that focuses on using cross-layer optimization for energy consumption is presented in \cite{Herrmann_energy_cross}. In this work, the authors demonstrate how cross-layer optimization can be implemented to extend the network lifetime. To this end, ISA 100.11 \cite{100_11_a} and a  Wireless highway addressable remote transducer (WiHART) \cite{WirelessHART} compatible sensor network is used in the context of petroleum refinery scenario. The authors demonstrate the utility of optimization that accounts for the fixed frame size constraint of current industrial wireless standards in smaller network scenarios. Their NS3 simulations depicted longer network lifetimes than the traditional approaches that use minimum hop routing for the industrial wireless network. At the same time, they do caution by stating that even frame-based optimization does experience a non-negligible failure rate when solving routes for large network sizes. 

A  mathematical model for cross-layer protocol optimization in SDR was proposed in \cite{Nosheen_cross_OMNET}. The goal of the framework is to provide tactical SDR with the flexibility to adapt its objectives including minimizing energy consumption while maintaining reliable packet delivery and latency constraints. Though the proposed solution is designed for SDRs the solution is only implemented and evaluated in OMNET++. Their simulations used a number of mission-critical network scenarios to demonstrate enhanced performance where SDRs effectively adapt to the dynamic environment.

As we survey through the breadth and width of cross-layer solutions applied to achieve various objectives discussed above, we can clearly see how the majority of the solutions rely on simulations that do not completely model or capture various intricacies of a real-world wireless network. While this is a necessary and valuable step to design, evaluate, and refine novel solutions, it often also ends up being the finish line for several of these approaches due to the daunting challenges that are involved in the maturation of the solutions. There have been attempts to overcome this challenge and shortcomings by designing frameworks, and utilizing SDR-based testbed. In the next section, we discuss some of these attempts that have striven to bridge the gap.

\subsection{Preliminary Testbed Evaluation}

In several cases, the next logical step is to evaluate the feasibility and performance on a hardware-based testbed. The advent of SDR has significantly nurtured efforts in this direction. At the same time, there is a distinction in the maturity/utility of solutions that have been implemented using a host PC controlling the SDRs and ones efficiently ported to a standalone e-SDR. The key difference is in the rapidity of development on computationally capable hardware in the first case as opposed to carefully optimized (often C/C++ or VHDL/Verilog) implementation on embedded (resource-constrained) hardware. The host PC-based development approach saves time and resources to rule out impractical solutions before significant time is spend in optimizing the implementation for final matured deployment. Therefore, based on the resources, the risk associated with novel cross-layer solutions either approaches can be adopted to mature network control and management solutions. To this end, several frameworks have been proposed for SDR to enable cross-layer optimized control \cite{shome_Cross_hardware, RcUBe, Combat}. There has also been work that employs SDR hardware but rely on emulation platforms \cite{soltani_cross_emulation} to perform evaluations as opposed to over-the-air (OTA) experimentations. Such emulation-based approaches may provide more flexibility to perform larger number of experiments in various topologies but still cannot be substituted for OTA evaluations or field trials. 

Several of the solutions discussed earlier have been successfully extended to preliminary hardware testbeds \cite{Sklivanitis15GLOBECOM,Jagannath18TMC}. In most of these cases, SDRs like the universal software radio peripheral (USRP) are used in association with the host PC. The authors extend their work of the proposed DRS algorithm \cite{DRS} by implementing it on a five-node USRP testbed \cite{Jagannath18TMC}. To accomplish this, the authors develop a cross-layer framework they refer to as CrOss-layer Based testbed with Analysis Tool (COmBAT) \cite{Combat} that is implemented in Python to run on host PC that controls the USRP SDRs. The authors compare the proposed solution with ROSA to demonstrate improvement in both effective throughput and reliability of the network. 

Similarly, in \cite{Sklivanitis15GLOBECOM} authors use four-node SDR testbed to demonstrate how cognitive channelization can be achieved by jointly optimizing the transmission power and the waveform channel of the secondary users. The experiments are conducted for both narrowband and wideband of primary users showing improvements in BER for both primary and secondary users.

A resource allocation approach for CRN was designed and implemented in \cite{Tsitseklis_cross_SDR} using two SDR-based testbeds of the ORCA federation \cite{ORCA}. The proposed distributed resource allocation employs a Markov Random Field (MRF) framework to be deployed on secondary nodes of the CRN. The implementation leveraged GNU Radio to implement functionalities such as spectrum sensing, collision detection, among others. The authors consider multi-channel CRNs and focus on the physical, data link layer (more specifically MAC), and network layer. The goal of any CRN is to ensure successful communication between secondary users in a dynamic environment without impacting the primary users. The secondary user SDRs are designed to calculate an energy function based on the current states of the neighbors. The objective of the secondary users is to minimize the local energy function which in turn minimized the interference. The authors claim that the secondary users asymptotically converge to global optimal solution by updating their energy function through local sampling. The authors successfully implement the proposed solution on two SDR testbeds, IRIS \cite{Iris} and ORBIT \cite{Orbit}. The experiments ranged from using up to 2 primary users and 9 secondary users. They compare their solution with reinforcement learning-based channel allocation algorithms. Through their preliminary experimental evaluation, they were able to demonstrate complete transparency towards primary users and demonstrate performance in terms of collision percentage and the number of required transmission slots for the secondary users of the CRN. It is interesting to note that the achievable throughput has not been discussed in this work which in most cases are throttled by lower sustainable sampling rates on the testbed. 

Beyond relying on SDRs, well-defined commercial wireless protocols like WiFi and LoRa (for physical layer) have also been used along with microcomputing platforms to design cross-layer approaches \cite{Jagannath19ADH_HELPER}. In this case, a cross-layer approach - distributed energy-efficient routing (SEEK) - was implemented on  Raspberry Pi to maximize the network lifetime using LoRa as the PHY. Using their proposed geographical routing protocol, SEEK, the authors were able to demonstrate significant improvement in the network lifetime compared to greedy ad hoc routing approaches. In this form, it was highly restricted in throughput (due to LoRa) as a trade-off for a longer transmission range. The survey of cross-layer approaches and their level of maturity has been summarized in Table \ref{table:SurveyTable}.

In all these examples, the achievable throughput is usually low due to the lower sampling rate or rely on the significant computing power of the host PC. Either of these factors renders such solutions highly restrictive in terms of utility, portability, fieldability, and often does not meet end-user requirements. In this article, we discuss how by customizing COTS low SWaP e-SDR with limited computing resources one can efficiently design and deploy the \emph{first-known embedded software-defined cross-layer optimized network}. 

\begin{table*}
\footnotesize
\caption{Summary table of cross-layer solutions and their maturity}
\begin{tcolorbox}[tab2,tabularx={|p{55pt}||p{50pt}|p{38pt}|p{38pt}|p{270pt}}]
\hline
Proposed Work & Optimizing/ \newline Framework & Simulation & Preliminary Testbed  &Comments \\
\hline \hline
Ding et al 2010 \cite{ROSA} & Throughput & \centering \checkmark &  &   Algorithm designed to perform joint routing, dynamic spectrum allocation, scheduling, and transmit power control for cognitive radio networks. \\\hline
Ding et al 2013 \cite{ROCH} & Throughput & \centering \checkmark &  &   Algorithm designed to perform routing and code division channelization in a CDMA based cognitive radio networks.\\\hline
Liu et al 2018 \cite{Liu_energy_throughput_cross} & Throughput & \centering \checkmark &  &   Throughput-maximal cross-layer framework is proposed for DSRC-based vehicular network. \\\hline
Jemili et al 2020 \cite{JEMILI_multipath} & Throughput & \centering \checkmark &  &  Cross-layer multi-path routing approach that utilizes non-correlated and node-disjoint paths \\\hline
Cheng et al 2013 \cite{Cheng_cross_Delay} & Delay & \centering \checkmark &  &    Two cross-layer approaches are proposed to minimize end-to-end delay for multihop wireless network and evaluated in simulations.\\\hline
She et al 2018 \cite{She_URLLC} & Reliability \newline Latency & \centering \checkmark &  &   The policies for packet dropping, power allocation, and bandwidth allocation are optimized to minimize the transmit power under the QoS constraint. \\\hline
Xlong et al 2011 \cite{Ekici_delay-aware} & Delay & \centering \checkmark &  &   A delay-aware joint flow control, routing, and scheduling algorithm is introduced for multihop network to maximize network utilization. \\\hline
Xue et al 2013 \cite{Ekici_delay-guaranteed} & Delay & \centering \checkmark &  &   Cross-layer approach that use virtual queue structures to guarantee finite buffer size or worst-case delay performance.\\\hline
Jagannath et al 2016 \cite{DRS} & Effective Throughput & \centering \checkmark &  &   Incorporates virtual queue length into the objective function of the cross-layer algorithm to mitigate the last packet problem.\\\hline
Wang et al 2006 \cite{Wang_cross} & Fairness & \centering \checkmark &  &   A dual-based algorithm and a penalty-based algorithm designed and evaluated to address the fair rate control problem in a multihop random access network. \\\hline
Park et al 2008 \cite{Park_cross_fairness} &  Fairness & \centering \checkmark &  &   Authors propose a cross-layer feedback mechanism to improve fairness among sending and receiving WiFi stations. The solution is implemented and evaluated in NS2.   \\\hline
Xu et al 2015 \cite{Xu_cross_fairness} &  Fairness & \centering \checkmark &  &   Cross-layer fairness-driven approach to improve video delivery while remaining fair to the competing TCP flows. The solution is implemented and evaluated in NS2.   \\\hline
Herrmann et al 2018 \cite{Herrmann_energy_cross} & Network Lifetime & \centering \checkmark &  &   Authors design cross-layer algorithm for extending lifetime of industrial network for petroleum refinery. The solution was evaluated using NS3. \\\hline
Salameh et al 2021 \cite{Salameh_cross_energy} & Energy consumption & \centering \checkmark &  &   Algorithm designed for CRN to minimize energy consumption while adhering to delay constraints and minimize impact on primary users. Algorithm outperforms non-cross-layer approaches in MATLAB simulations. \\\hline
Jemili et al 2021 \cite{JEMILI_energy} & Energy Consumption & \centering \checkmark &  &   Cross-layer multi-path routing approach designed to support duty-cycled WSN networks for energy efficiency while reducing impact on response time delays. \\\hline
Lahane et al 2021 \cite{Lahane2021SecuredCC} & Energy Consumption & \centering \checkmark &  &   Cross-layer clustering based routing protocol that was shown to outperform traditional approaches like grouped grey wolf search optimization. \\\hline
Al-Jemeli et al 2015 \cite{Al-Jemeli_energy_cross} & Energy Consumption & \centering \checkmark &  &   Cross-layer network operation model proposed for energy efficiency in IEEE 802.15.4-based networks. The solution was evaluated using NS2. \\\hline
Nosheed et al 2019 \cite{Nosheen_cross_OMNET} & Energy efficiency & \centering \checkmark &  &   Mathematical model designed for SDR and performance evaluated in OMNET++. \\\hline
Herrmann et al 2018 \cite{Herrmann_energy_cross} & Network Lifetime & \centering \checkmark &  &   Authors design cross-layer algorithm for extending lifetime of industrial network for petroleum refinery. The solution was evaluated using NS3. \\\hline
Dai et al 2018 \cite{Dai_energy_cross} & Energy efficiency & \centering \checkmark &  &   Authors aim to maximize energy efficiency by jointly optimizing modulation, transmit power, and route. \\\hline
 \hline
Shome et al 2015 \cite{shome_Cross_hardware} & Framework &   & \centering \checkmark   & Framework for implementation and evaluation of cross-layer approaches.\\\hline
Demirors et al 2015 \cite{RcUBe} & Framework &   & \centering \checkmark  & Framework for implementing and evaluating cross-layer approaches.\\\hline
Jagannath et al 2016 \cite{Combat} & Framework &   & \centering \checkmark   & Framework for implementing and evaluating cross-layer approaches.\\\hline
Soltani et al 2019 \cite{soltani_cross_emulation} & Not specified &   & \centering \checkmark   &  Designed to integrate dynamic spectrum access, backpressure algorithm, and network coding for multihop networking. Evaluation performed using emulations.\\\hline
Jagannath et al 2018 \cite{Jagannath18TMC} & Throughput & & \centering \checkmark   & Throughput and reliability performance evaluated on 5-node USRP N210 SDR testbed.\\\hline
Sklivanitis et al 2015 \cite{Sklivanitis15GLOBECOM} & SINR &  & \centering \checkmark   & BER performance evaluated on 4-node USRP N210 SDR testbed.\\\hline
Tsitseklis et al 2020 \cite{Tsitseklis_cross_SDR} & Avoid Collision & &  \centering \checkmark &   Cross-layer framework designed for CRN and evaluated on 2-node SDR testbed named IRIS \cite{Iris} and ORBIT \cite{Orbit}. \\\hline
SEEK 2019 \cite{Jagannath19ADH_HELPER} & Network Lifetime  &  & \centering \checkmark  & Algorithm considers link energy efficiency, backlog, forward progress, \& residual energy. Solution is implemented on RPI \& evaluated on 6-node testbed.\\\hline
\end{tcolorbox}{}
\label{table:SurveyTable}
\end{table*}

In concluding this section, we have seen that the majority of the cross-layer optimized works are limited to simulations and have not been successfully validated on a hardware platform. This is a major hurdle and shortcoming of the current state of research and development. In the recent past, some of this has been mitigated by preliminary hardware-based evaluations but has often provided limited performance in terms of metrics like throughput (due to low sampling rate constrained by computations) and/or has been dependent on external computational platforms. Due to these reasons, even with the advances made in the field, there is no known cross-layer optimized e-SDR solution built using commercial-off-the-shelf (COTS) e-SDR that, (i) can be deployed as a standalone unit, (ii) makes cross-layer optimized distributed routing decisions, (iii) is ruggedized for outdoor deployment, and (iv) can provide reliable and high throughput (up to 11 Mbps) links over large distances (1 km for up to 5.5 Mbps). The difficulty of finding the right trade-off between software design choices and performance along with achieving standalone, fieldable hardware customization is a daunting task. This article provides such a solution for the first time showing that designing and deploying high Technology Readiness Level (TRL) cross-layer optimized solutions based on a COTS low SWaP e-SDR is feasible.

\section{System Design}
\label{Sec:SystemDesign}

In this section, we first discuss some core design challenges that may be encountered during similar endeavors. Thereafter, we describe the system design of the cross-layer optimized transceiver and how it has been executed.

\subsection{Design Challenges}

\subsubsection{Choice of COTS SDR hardware} In recent years, there has been a surge in the number of SDRs available in the market. While this has provided several tradeoff opportunities for system design decisions, yet, determining the appropriate choice is challenging due to the multi-dimensional tradeoff that exists with the decision making. At a high level, these choices can be divided into some core characteristics portability (size and weight), embedded computational capabilities, RF frontend parameters (instantaneous bandwidth, tunable frequency range, among others), and cost.

In our case, portability and embedded computational capability were the two prime candidates. This meant ruling out all the SDRs that may not have sufficient embedded computational resources. In this article, we refer to SDR with embedded processing capability (such as Field Programmable Gate Array (FPGA) and General Purpose Processors (GPP)) as e-SDR. Generally, the larger the SDR the higher the computational resources but may end up consuming higher power and leading to a larger design. Hence, at this stage, the designer has to determine the best tradeoff that works for a given design and make sure the constraints from this point onwards is acceptable to the overall software architecture.

\subsubsection{Limited computational resources} 

Embedded radio programming is a non-trivial task and the primary factor that must be considered are the available computational resources. Accordingly, care must be taken to avoid redundant memory read and write operations to reduce energy consumption and latency of operation. From a protocol stack implementation perspective, the available accelerators are the CPU and FPGA housed in the Xilinx Zynq XC7Z010-2I SoC with 512 MB of DDR3L RAM and 128 MB of QSPI Flash memory on the radio platform. These impose constraints on the number and size of the packet queues as well as storage of configuration and other header information on the radio.

Rather than going for a top-down approach whereby the protocol stack is designed and simulated with evaluations followed by hardware implementation, we chose a bottom-up approach where these radio resource constraints are carefully considered right from the inception of the protocol stack. These are evident from the energy consumption factor considered in the routing, design of a segment size where the packets are grouped into segments for outbound transmission, and gathering Optimization Assisting Information (OAI) via piggybacking. This way the control packet exchange won't occur per packet rather for a segment of packets. These steps are adopted to reduce overhead which implicitly reduces latency and energy consumption. Hence, the hardware constraints, as well as the target application requirements (long-term unattended deployment), directed our design choices.

Most of the FPGA resources were utilized for the PHY layer which is a pure IEEE802.11b standard. The PHY was housed in the FPGA for reduced latency operation. Since the FPGA resources were nearly fully utilized for the 802.11b implementation, none of the remaining upper layers could be offloaded to FPGA for acceleration. Consequently, the entire stack except the PHY/L1 was implemented on the Dual-core ARM Cortex A9 CPU of the SoC. The software architecture utilizes both the user space as well as kernel space. To reduce repeated I/O operations and to parallelize the stack operations to a full extent, we leveraged daemons executing from the user space that hosted multi-threaded processes. A deep dive into the software implementation is elaborated in section \ref{sec:xl}. The software architecture is designed and developed to execute agnostic to the underlying PHY/L1 layer, accordingly, we have followed a plug-and-play design ethos for the software stack. Our implementation demonstrates this by interfacing the software protocol stack executed fully from the ARM processor with the 802.11b FPGA core.

\subsubsection{Gathering optimization assisting information}
Network optimization can be generally divided into two categories, (i) centralized and (ii) distributed. Centralized approaches often provide globally optimal solutions but at the expense of large overhead incurred in accumulating global information. This also introduces delays and would suffer scalability issues for larger network sizes. In contrast, distributed approaches reduce the overhead and are in general scalable but may not converge to globally optimal solutions. In either case, there is a need to acquire information that is essential to execute cross-layer optimization.

For ad hoc networks that are highly dynamic and need to be scalable, distributed approaches are the desired choice. This is also the case for our solutions. Our approach (which will be discussed in detail in the next few sections) rely on gathering OAI from immediate neighbors. To reduce the overhead, this information can be appended to the control packets such as RTS, CTS, BEACON, etc. The frequency of BEACON packets can be a function of network parameters such as relative mobility of the nodes, rate of change of traffic, among others. The essential takeaway is the need to carefully design protocols to ensure efficient yet effective exchange of timely information for the cross-layer decision engine. 

\subsection{Embedded Software Defined Radio Platform}

\begin{figure}[h!]
\centering
\includegraphics[angle=90, width=0.9 \columnwidth]{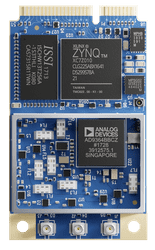} 
\caption{Target baseline Z2 e-SDR}
\label{fig:Z2}
\end{figure}

One of the key objectives of the work was to develop a modular, programmable, portable, handheld, battery-powered, standalone solution which should operate in harsh conditions for several hours. This implied that the foundation of the design needs to be a low SWaP e-SDR. The proposed solution was implemented on a Epiq Solutions' Sidekiq Z2 e-SDR \cite{sidekiqz2} (Figure \ref{fig:Z2}). It consists of an Analog Devices' AD9364 RFIC, Xilinx Zynq XC7Z010-2I system on chip (SoC), and the key device specifications are provided in Table \ref{tab:z2}.

\begin{table}[h]
 \centering
 \captionof{table}{Specifications of Z2}
    \begin{tcolorbox}[tab2,tabularx={|p{3 cm}||p{5 cm}}]
      \textbf{Specs.}   & \textbf{Values}  \\ \hline\hline
      Frequency   & 70 MHz - 6 GHz  \\ \hline
    Sample Rate & Up to 61.44 MS/s   \\ \hline
    RAM & Up to 512 MB of DDR3L   \\ \hline
    Flash memory &  128 MB  \\ \hline
    Temperature rating & $-40 C$ to $+85\deg C$ \\ \hline
    Processor & Dual-core ARM Cortex A9   \\ \hline
    Size & 30 x 51 x 5 mm \\ \hline
    Weight & 8 grams \\ \hline
    \end{tcolorbox}{}
    \label{tab:z2}
\end{table}

\begin{figure*}[h!]
\begin{minipage}[h]{0.4 \linewidth}
\centering
\includegraphics[width=.99 \columnwidth]{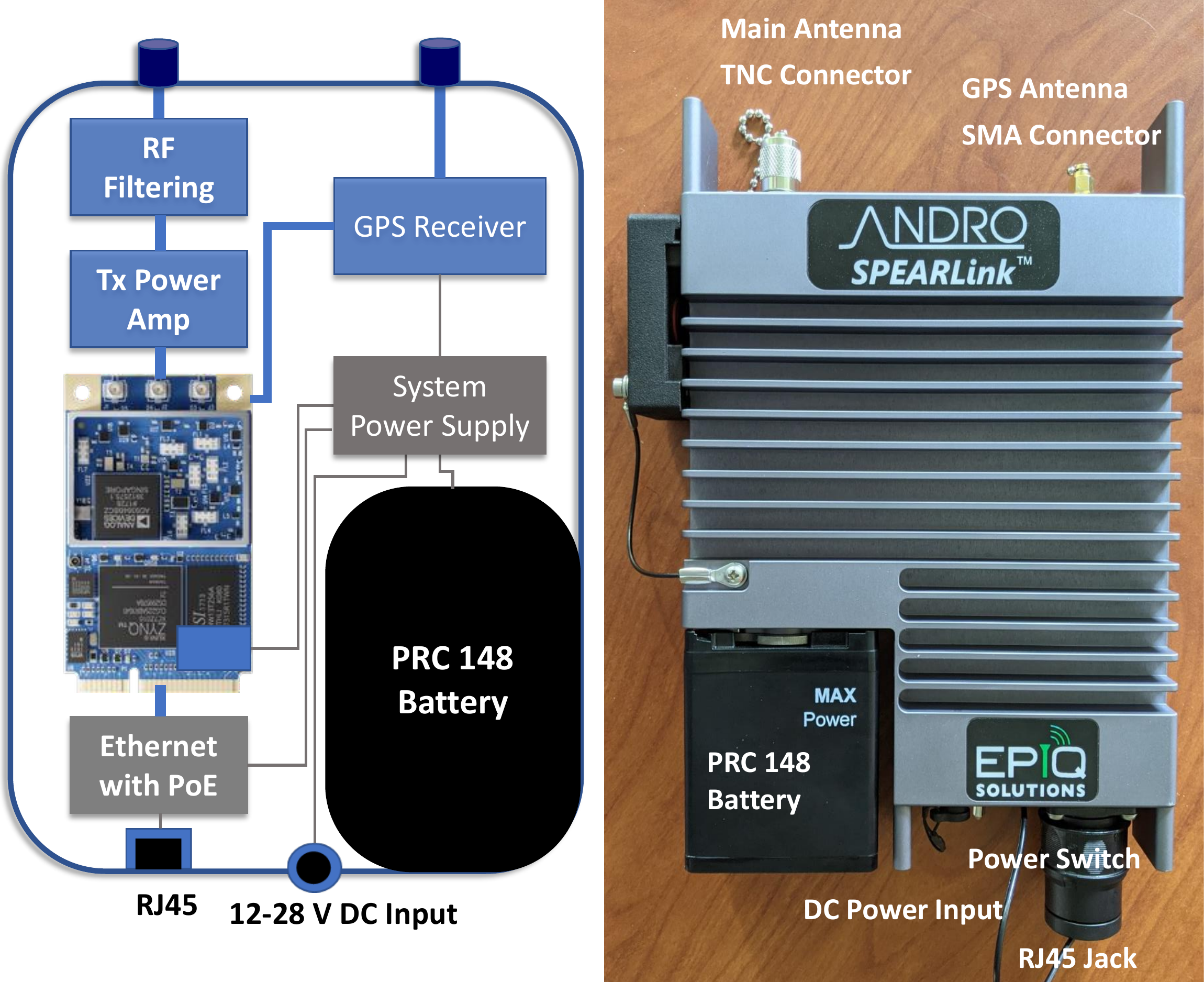} 
\caption{Transceiver Design}
\label{fig:Spearlink}
\end{minipage}
\begin{minipage}[h]{0.6 \linewidth}
\centering
\includegraphics[width=.95 \columnwidth]{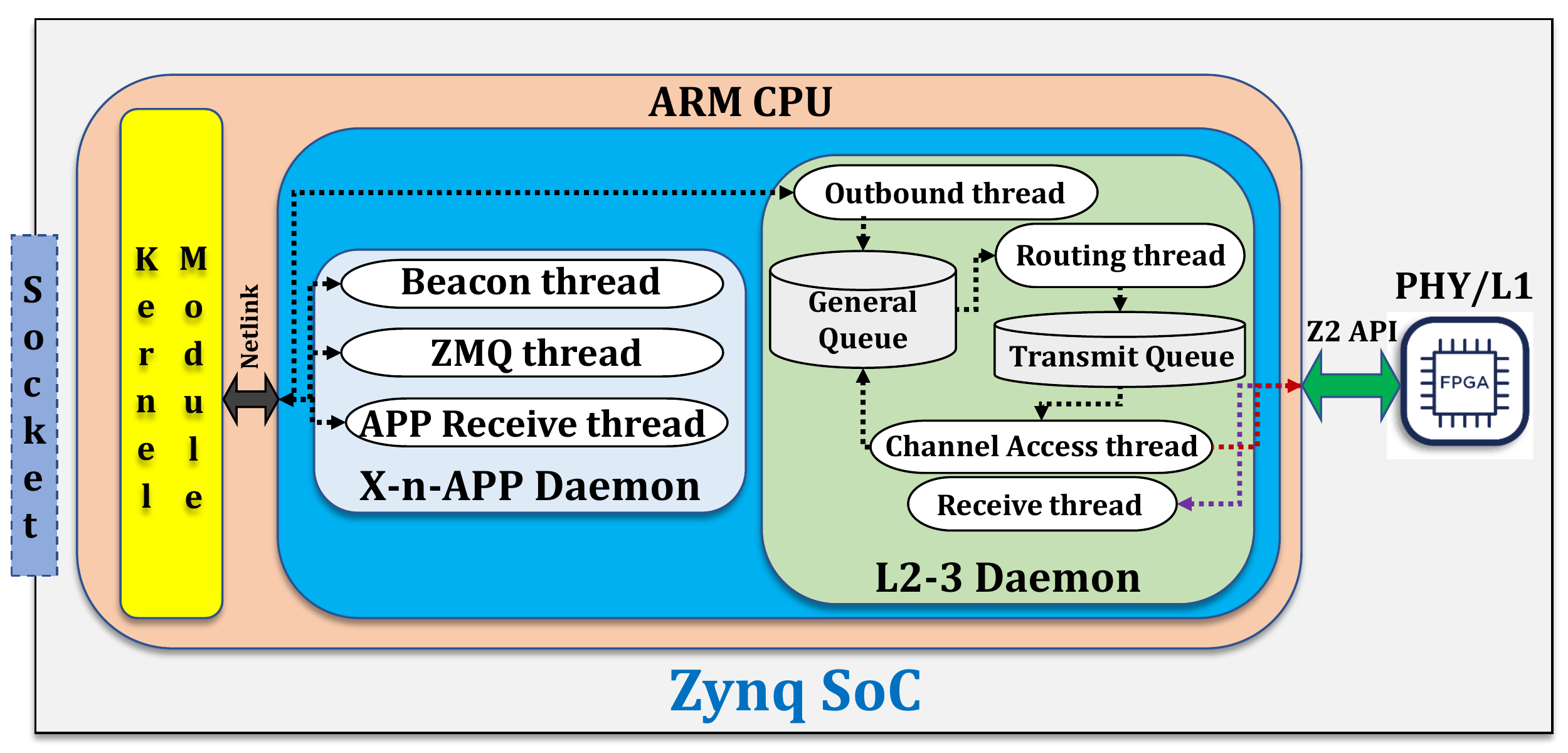} 
\caption{Software Architecture.}
\label{fig:SW}
\end{minipage}
\end{figure*}

Several hardware customizations were necessary to accomplish the objective of designing a fieldable transceiver using the COTS e-SDR. This includes adding a power amplifier, filters for the frequency of interest, power supply system that is capable of supplying power from the battery during standalone remote operation but could also operate from a direct current (DC) power source when available. To aid the implementation of various cross-layer routing techniques that use the location of nodes (such as geographical routing \cite{Jagannath19ADH_HELPER}), an embedded GPS receiver was also included in the final design. The block diagram of the final transceiver and the ruggedized prototype is shown in Figure \ref{fig:Spearlink}.

\subsection{Cross-layer Software Architecture for Embedded System}

\label{sec:xl}

\subsubsection{Design Goals and Overall Architecture}

The custom cross-layer protocol stack for the transceivers is implemented on the Zynq SoC of the Sidekiq Z2 platform. Specifically, the PHY is an IEEE802.11b implemented purely on the FPGA subsystem while the upper layers are implemented in the C/C++ language on the embedded Linux operating system of the Dual-core ARM Cortex A9 CPU of the SoC. 

As mentioned previously, our design principle is to develop a plug-and-play software module of the protocol stack (except L1) to interface with any underlying PHY layer. The 802.11b FPGA core is a direct sequence spread spectrum (DSSS)-Complementary code keying (CCK) PHY with a 22MHz standard channel bandwidth which nearly fully utilized the logic cells and DSP blocks of the FPGA. The supported data rates with DSSS are 1 Mbps and 2 Mbps while CCK modulation supports 5.5 Mbps and 11 Mbps. The software stack is designed to be agnostic to the PHY layer and merely interfaces with the PHY FPGA core over a C-level SKIQ80211 application programming interface (API) provided by Epiq Solutions. The API provides all the necessary function calls to utilize the RF frontend.

We would like to reemphasize that the solution can be adapted to use most if not all PHY layer standards as long as the performance parameters (bit error rate, transmit power, data rates, etc.) are communicated to the upper layers over an appropriate C/C++ level interface. Since the focus of this work is the software implementation and evaluation of the upper layers (above PHY/L1), in the remainder of the section, we will detail the software architecture of the upper layers to facilitate long-range mesh networking. 

Our primary goal in building the software architecture was to maintain \emph{reconfigurability}. The reconfigurability was enforced by adopting a modular design framework with defined functions for each module while being resource-efficient. We define reconfigurability as the ability to modify the protocol characteristics such as routing objective, specifics of the cross-layer information exchanged, etc. Accommodating a software-defined cross-layer protocol stack in an ARM CPU is a non-trivial challenge. The software-defined architecture reaps benefits from the flexible software architecture which eases the future upgradability, customization (of factors like optimization objective among others), and maintenance of the transceiver software. 
Essentially, designing the entire radio stack on the SoC presents a daunting challenge owing to the memory, computational, and latency constraints. An unorganized framework could add overhead from unnecessary resource utilization consequently increasing the system latency. 

The software architecture is broadly categorized into user and kernel space with daemons running in the user-space as in Figure \ref{fig:SW}. We resort to daemons each of which hosts its own threads to attain a parallelized architecture that does not perform redundant and unnecessary memory accesses. In other words, we can say the user-space hosts multi-threaded processes which interface with the kernel space. However, there are certain design challenges with moving key functions to the kernel space. Kernel programming requires the most trusted operations as any bug or corruption may cause severe system crashes. Nevertheless, kernel space enjoys the benefits of low latency memory access operations. Furthermore, we also leverage the transport layer and IP headers supported by the Linux kernel. The software architecture in Figure \ref{fig:SW} shows a \textit{socket} which is a generalized custom socket architecture for interfacing with external devices. The socket architecture is generalized to be able to reprogram and enable multiple interfacing options with a wide range of socket protocols such as ZMQ, UDP, etc.

The stack employs two daemons, namely; L2-3 and Cross-layer \& Application (X-n-APP) that interact with the kernel module via Netlink sockets. Notice that the entire stack only has two daemons since it's more efficient to use threads over processes as they share the process' resources. On the other hand, having more processes would require more computational and memory resources. The kernel module handles the transport layer, packet encapsulation, and partial IP header population prior to passing over to the L2-3 daemon. The L2-3 daemon is the cross-layer L1, L2 (MAC), and L3 (Network) module which performs CSMA/CA-based medium access and cross-layer routing. Another level of cross-layer interactions occurs when the L2-3 daemon acquires L1 information such as the link reliability, data rate, etc., via the Sidekiq Z2's SKIQ80211 API. Hence, the term cross-layer as it involves interaction between L1, L2, and L3 to perform the optimized decision making. 

\subsubsection{Cross-layer Routing Protocol Overview}

The desired use case application of this work was to deploy an independent energy-aware ad hoc network in large remote areas where traditional communication infrastructure for networks such as 5G, WiFi, or LTE are not readily available. A prime example of such an application are large test and evaluation (T\&E) ranges that are usually set up in remote places, collecting data from networks of remote installations, supporting rescue operations after a disaster where traditional communication infrastructure is unavailable, among others. In many of these cases the core requirements of a desired routing protocol are as follows,

\begin{itemize}
\item The solution must be easy to deploy in ad hoc manner such that nodes can move in and out of the network without effecting the overall network performance.
\item The solutions should be able to operate in a distributed manner such that it can scale without incurring large overhead.
\item Operates in an energy efficient manner to maximize network lifetime since it is challenging to change batteries regularly in a large remotely situated network.
\item Solutions should be able to adapt to changing network traffic patterns and avoid congested paths by load balancing.
\end{itemize}

Keeping the above requirements in mind, the \emph{diStributed Energy Efficient bacKpressure  (SEEK)} routing algorithm \cite{Jagannath19CCNC_1, Jagannath19ADH_HELPER} is leveraged in this work which utilizes the geographic information of nodes, differential queue backlog, residual battery energy, and link reliability to compute the optimal next hop. In this section, we will present a formal derivation of our utility function ($\mathcal{U}_{ij}$ with respect to wireless link $\mathfrak{L}_(i,j)$) between node $i$ and node $j$ and formulate the network optimization problem. 

The utility function considers the following parameters associated with potential next-hop; (i) proximity to destination, (ii) differential queue backlog, (iii) residual battery energy, (iv) energy efficiency of the link, and (v) the corresponding link throughput. This information is gathered from traditional control packets like the Beacon packets. The Beacon packets will contain updated OAI. The energy efficiency of a given link can be expressed as follows \cite{WBAN},

\begin{equation}
\mathcal{\eta}_{ij}= \frac{\mathrm{R}_{b}^{ij} \left(1 - \mathrm{e}_{b}^{ij}\right)}{\mathtt{P}_{ij}} = \frac{\mathrm{G}_{ij}}{\mathtt{P}_{ij}}
\end{equation}

where $\mathcal{\eta}_{ij}$ gives the measure of the number of bits successfully transmitted over $\mathfrak{L}_(i,j)$ per Joule of transmission energy. Here, $\mathrm{G}_{ij}$ is the corresponding goodput measure and $\mathrm{e}_{b}^{ij}$ is the bit error rate with respect to $\mathfrak{L}_(i,j)$, $\mathrm{T}_{ij}$ is the transmission strategy which includes choice of transmission data rate $\mathrm{R}_{b}^{ij}$ and transmit power $\mathtt{P}_{ij}$.

Another key factor that needs to be considered in routing is the differential queue backlog ($\Delta\mathcal{Q}_{ij} = \mathtt{q}_{i} - \mathtt{q}_{j}$) with respect to the source node ($i$) and next-hop ($j$) \cite{backpressure, Wt_backpressure, DRS}. The queue backlog at the destination node is considered to be zero. Considering the queue backlog is necessary to mitigate congestion in the network and traditional backpressure algorithms have been shown to be throughput optimal \cite{backpressure}. Since achieving maximum throughput is not the sole objective of SEEK algorithm, the differential backlog is just one parameter in our utility function. The effective progress made by a packet can be represented as $d_{is}-d_{js}$. Choosing nodes that provide larger progress implies fewer hops to the sink node which in turn could lead to smaller energy consumption. Finally, to ensure uniform depletion of energy per node, we need to consider the $\mathtt{E}^{j}_{r}$ (residual energy) of potential next hops \cite{GPNC}. Therefore, we define our utility function as follows,

\begin{equation}
    \mathcal{U}_{ij} = \mathcal{\eta}_{ij}  \overbracket[0.8pt]{\left(\frac{\max\left[\Delta\mathcal{Q}_{ij},0 \right]}{\mathtt{q}_{i}}\right)}^\text{\clap{Differential Backlog~}} \underbracket[0.8pt]{\left(\frac{d_{is}-d_{js}}{d_{is}} \right)}_\text{\clap{Effective Forward Progress~}} \overbracket[0.8pt]{\left( \frac{\mathtt{E}^{j}_{r}}{\mathtt{E}^{j}_{0}} \right)}^\text{\clap{Relative Residual Energy~}}, \forall j\in \mathbb{NB}_{i} 
    \end{equation} 


$\eta_{ij}$ aims to improve the energy efficiency of the network and can be replaced by reliability when using constant power and modulation. It is also interesting to note that the maximum value of $ \mathcal{U}_{ij}=\mathcal{\eta}_{ij}$ when each of the three normalized terms is $1$. This implies that each of the other terms penalizes the utility function based on the instantaneous value. For example, a small differential backlog ($\mathtt{q}_{i} - \mathtt{q}_{j}<\mathtt{q}_{i}$) will dampen the value of $\mathcal{U}_{ij}$. Both $d_{is}-d_{js}$ and $\mathtt{E}^{j}_{r}$ will have similar effects on $\mathcal{U}_{ij}$.  

The objective of the network is to maximize the summation of $\mathcal{U}_{ij}$ for all possible links $\mathfrak{L}_(i,j)$ in order to maximize the overall energy efficiency of the network. This, in turn, will ensure reliable communication while maximizing the network lifetime (which is defined as the time when the first node in the network depletes its energy leading to a network hole). The optimization problem is subject to residual battery energy, queue backlog, bit error rate, and capacity constraints. This is formulated as Problem $\mathcal{P}_1$ shown below,

\begin{align}
\mathcal{P}_1\!:\textup{Given}\!&: \mathcal{G}(\mathbb{N}_{net},\mathbb{E}),\;\;\mathbb{G},\;\; \mathbb{E}_{r},\;\; \mathbb{Q}\notag \\
				\textup{Find}\!&: \mathbb{NH}^{*}, \mathbb{T}^{*}\notag \\
				\textup{Maximize}\!&: \sum_{i \in \mathbb{N}_{net}} \sum_{j \in \mathbb{NB}_{i}} \mathcal{U}_{ij}\\
				\textup{subject\; to}\!&:\notag \\
				& \mathrm{R}_{b}^{ij} \leq C_{ij}, \;\;\;\;\; \forall i \in \mathbb{N}_{net},\; \forall j \in \mathbb{NB}_{i} \label{constr:capacity} \\ 
                &\mathrm{e}_{b}^{ij} > \mathrm{e}_{b*}^{ij}, \;\;\;\;\;\;\; \forall i \in \mathbb{N}_{net},\; \forall j \in \mathbb{NB}_{i} \label{constr:error} \\ 
				& \mathtt{E}^{i}_{r} > 0,\;\;\;\;\;\;\;\;\;\; \forall i \in \mathbb{N}_{net} \label{constr:Energy} \\
                & \mathtt{q}_{i} \geq 0,\;\;\;\;\;\;\;\;\;\; \forall i \in \mathbb{N}_{net} \label{constr:backlog} 
\end{align}

where the objective is to find the set of next-hop and transmission strategy for all nodes in the network which can be represented as $\mathbb{NH}^{*} = [\mathrm{NH}_{i}^{*}]$ and $\mathbb{T}^{*} = [\mathrm{T}_{ij}^{*}]$ respectively, $\forall i \in \mathbb{N}_{net} \;j \in \mathbb{NB}_{i}$. In the above optimization problem $\mathcal{P}_1$, $\mathbb{G} = [\mathrm{G}_{ij}]$, $\mathbb{E}_{r} = [\mathtt{E}^{i}_{r}]$ and $\mathbb{Q} = [\mathtt{q}_{i}],\; \forall i \in \mathbb{N}_{net},\; \forall j \in \mathbb{NB}_{i}$ denote the set of goodput measure, residual battery energy and queue backlogs respectively. The constraint \ref{constr:capacity} restricts the total amount of data rate in link $(i, j)$ to be lower than or equal to the physical link capacity. Constraint \ref{constr:error} imposeS that any transmission should guarantee the required BER. Finally, constraints \ref{constr:Energy} and \ref{constr:backlog} ensure the residual energy and queue backlog of each node will not have negative values. It can be seen that for solving the above optimization problem, nodes would require global knowledge of the network. This would require the centralized controller to maintain largeR network tables storing node parameters which will eventually flood with data as the the network grows in size.
Since the centralized optimization method is not a scalable solution, this motivates the need for a scalable distributed solution. We adopt SEEK which will operate in a distributed fashion and enable each node to find the next-hop based on the local information available to them. Each node with a packet to transmit chooses an optimal next-hop and transmission parameters such that it maximizes its own local utility function. This can be considered as a divide-and-conquer approach to solving the optimization problem in a distributed manner. Accordingly, every source node ($i$) will aim to maximize the utility function $\mathcal{U}_{ij}$ and select the optimal next-hop and transmission strategy as follows,
\begin{equation}
[j^*,\mathrm{T}_{ij}^*] = \arg\max_{j}\: \mathcal{U}_{ij}, \forall j\in \mathbb{NB}_{i} 
\end{equation}

\begin{figure}
    \centering
    \includegraphics[width=1 \columnwidth]{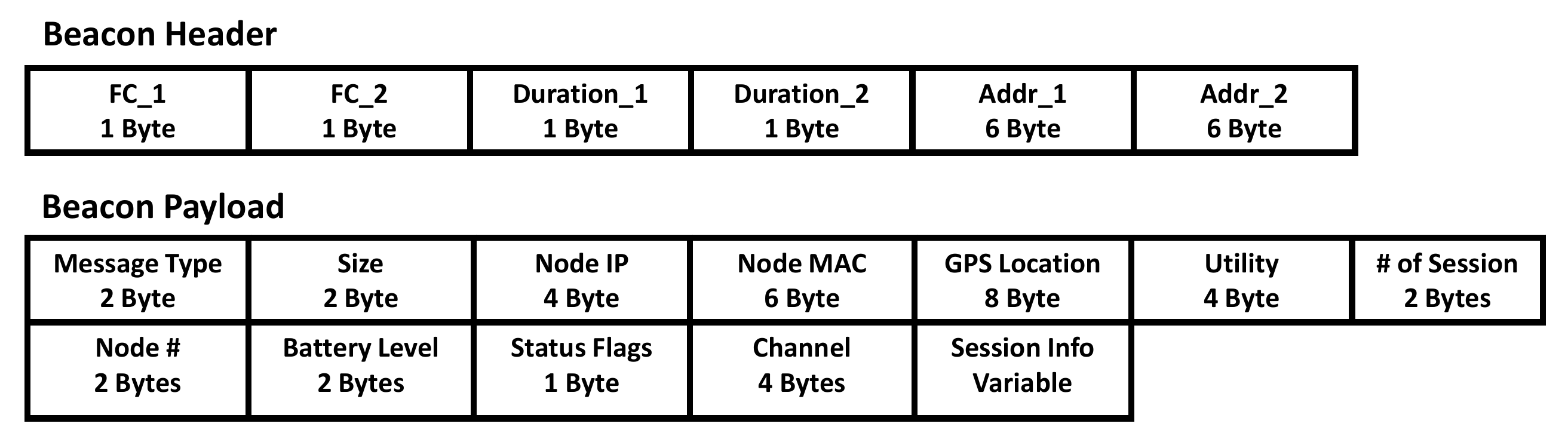}
    \caption{Beacon packet structure}
    \label{fig:beacon}
\end{figure}

Each node will maintain a neighbor table with node parameters of its neighbors and will update the table as needed based on information from the beacon packets. The structure of the beacon packet is depicted in Figure \ref{fig:beacon} with self-explanatory field names. The effectiveness of SEEK has been demonstrated in \cite{Jagannath19ADH_HELPER}. The emphasis of this work and contribution is not SEEK but the design and feasibility of maturing similar solutions for tactical applications. \textit{Due to the modular implementation, just by changing a few lines of code that defines the utility function, one can reconfigure the stack to execute a new optimization objective.} This flexibility in adapting the routing metric is demonstrated in the code snippet in Listing \ref{lst:route}.

\begin{lstlisting}[language=C++, caption=Routing Utility Computation Snippet - Battery powered scenario,label={lst:route}]
double calculateUtility(neighborThroughputInfo& sourceNode,neighborThroughputInfo& destinationNode)
{
    // Compute reliability term
    double reliability = reliabilityMean();
    
    // Compute forward progress term
    double distanceToTarget=haversine_distance(sourceNode.latitude,sourceNode.longitude,destinationNode.latitude,destinationNode.longitude);
    double distanceToTargetFromNextHop =haversine_distance(latitude,longitude,destinationNode.latitude,destinationNode.longitude);
    double forwardProgress =((distanceToTarget-distanceToTargetFromNextHop)/distanceToTarget);
     
    // Compute buffer backlog term           
    double sourceBacklog=(double)sourceNode.backlog;
    if(sourceBacklog==0) 
        sourceBacklog=1;
    
    double nextHopBacklog = (double)backlog;
    if(isInfoPotentiallyStale()) 
        nextHopBacklog=0;
    
    double minimumValue=0;
    double diffBacklog=0;
    if(nextHopBacklog>0)
        minimumValue=1/nextHopBacklog;
    double BacklogTerm=(sourceBacklog-nextHopBacklog)/sourceBacklog;
    if(nextHopBacklog<sourceBacklog) 
        diffBacklog=BacklogTerm;
    else 
        diffBacklog=std::max(BacklogTerm,minimumValue);
        
    // Compute residual battery energy
    double nextHopBattery = (double)battery;
    double residualBattery = (nextHopBattery)/100;
    
    // Compute utility term - Routing metric
    double routeUtility=reliability*forwardProgress*diffBacklog*residualBattery;
    
    return routeUtility;}
\end{lstlisting}
Here, line 35 shows the routing utility metric computation. This metric can be flexibly changed to include other route optimization metrics. For example, if all the nodes are powered by direct DC power supply and not on battery, then the routing metric can be modified to address this scenario as in Listing \ref{lst:route2}.

\begin{lstlisting}[language=C++, caption=Routing Utility Metric Modification in scenario when all nodes have DC power supply,,label={lst:route2}]
// Compute utility term - Routing metric
double routeUtility=reliability*forwardProgress*diffBacklog;
\end{lstlisting}

Finally, for the benefit of the readers, we also provide Table \ref{tab:notations} summarizing all the notations/symbols used in the discussion above. 

\begin{table}[h]
 \centering
 \captionof{table}{Summary of Notations}
    \begin{tcolorbox}[tab2,tabularx={|p{1.2 cm}||p{6.5 cm}}]
      \textbf{Symbols}   & \textbf{What it represents}  \\ \hline\hline
      $\mathfrak{L}(i,j)$   & Wireless link between node $i$ and node $j$   \\ \hline
      $\mathcal{U}_{ij}$   & Cross-layer utility function representing the efficacy of wireless link $\mathfrak{L}(i, j)$   \\ \hline
    $\eta_{ij}$ & Number of bits successfully transmitted over $\mathcal{L}(i, j)$ per Joule of transmission energy   \\ \hline
    $\mathrm{G}_{ij}$ & Goodput of wireless link $\mathfrak{L}(i, j)$    \\ \hline
    $\mathrm{e}_{b}^{ij}$ & Bit error rate of wireless link $\mathfrak{L}(i, j)$\\ \hline
    $\mathrm{e}_{b*}^{ij}$ & Minimum acceptable bit error rate \\ \hline
    $\mathrm{R}^{ij}_b$ & Data rate chosen for $\mathfrak{L}(i, j)$  \\ \hline
    $\mathtt{P}_{ij}$ & Transmit power chosen for $\mathfrak{L}(i, j)$  \\ \hline
    $\mathrm{T}_{ij}$ & Transmission strategy comprising of data rate and transmit power  \\ \hline
    $\mathtt{q}_{i}$ & Backlog of node $i$ \\ \hline
    $\Delta Q_{ij}$ & Differential Backlog \\ \hline
    $d_{is}$ & Distance between node $i$ and destination $s$   \\ \hline
    $\mathtt{E}^{j}_{r}$ & Residual energy of node $j$ \\ \hline
    $\mathtt{E}^{j}_{0}$ & Total initial energy of node $j$ \\ \hline
    $\mathbb{NB}_{i}$ & Set of neighbors of node $i$ \\ \hline
    $\mathcal{P}_1$ & Optimization problem 1 \\ \hline
    $\mathcal{G}(N,E)$ & Graph with nodes $N$ and edges $E$ \\ \hline
    $\mathbb{N}_{net}$ & Set of all nodes in the wireless network \\ \hline
    $\mathbb{E}$ & Set of all edges between nodes in the network \\ \hline
    $\mathbb{E}_{r}$ & Set of residual energies of all nodes in the network \\ \hline
    $\mathbb{G}$ & Set of all goodput measures \\ \hline
    $\mathbb{Q}$ & Set of all node backlogs \\ \hline
    $\mathbb{NH}^{*}$ & Set of all optimal next hop for the given time step  \\ \hline
    $\mathbb{T}^{*}$ & Set of all optimal transmission paramters for the given time step \\ \hline
    $\mathbb{NH}_i$ & Set of all neighbors of node $i$ \\ \hline
    $C_{ij}$ & Maximum channel capacity \\ \hline
    \end{tcolorbox}{}
    \label{tab:notations}
\end{table}

\subsubsection{L2-3 Daemon}
The outbound DATA packets that are handed over to the L2-3 daemon from the kernel module are queued in the \textit{General Queue} awaiting the best route assignment. This is accomplished by the \textit{Outbound} thread which continuously listens for incoming packets from the kernel module. The \textit{SEEK} thread continuously performs route computation and assignment for the outbound packets in the \textit{General Queue}. Following route assignment, the packets will await their transmission opportunity in the \textit{Transmit Queue}. A third thread - \textit{Channel Access} - performs CSMA/CA awaiting a clear-to-send (CTS) from its intended next-hop/destination to dequeue the segment (set of DATA packets) from the \textit{Transmit Queue}. The segment for which the CTS was received will be forwarded to the FPGA via the L1 API for \textit{over-the-air} transmission. It must be noted that if the intended next-hop for a segment is unresponsive, the segment will be returned back to the \textit{General Queue} for rerouting. Additionally, a \textit{Receive} thread continuously monitors for incoming packets from the FPGA, processes, and responds (such as with CTS or ACK) accordingly based on received packet type. A supplementary auxiliary thread also tracks DATA timeouts. 

\subsubsection{X-n-APP Deamon}

The X-n-APP daemon handles two tasks; (i) preparing the beacon packet with the necessary information to be shared with immediate neighbors for distributed optimization, and (ii) handling packets for any application such as a graphical user interface (GUI) in this case. 
The beacon packets as well as the GUI requires location information to reflect the most recent coordinates on the GUI. We choose an efficient approach and access the GPS module from only one location - X-n-APP daemon - in the stack to avoid GPS pinging from multiple software locations. This retrieved GPS location information is subsequently populated in the \textit{Beacon} packet. The GPS location extraction as well as beacon packet construction is carried out by the \textit{Beacon} thread. The \textit{Beacon} thread periodically sends these beacon packets to the lower layers for outbound transmission where the remaining fields such as residual battery and current buffer backlog are updated. Here, we note that, unlike the DATA, the beacon packet is treated as a control packet and is not queued in the general or transmit queues of the L2-3 daemon rather it is directly sent to the L1 FPGA for transmission.

The X-n-APP daemon also functions as a generalized application daemon for direct interfacing with an application such as a GUI. Both \textit{ZMQ} and \textit{APP receive} threads are designed to support such applications and hence can be customized based on the application at hand. Since the use case of this particular transceiver is remote deployment in test and evaluation ranges, a GUI application is developed to interface with the transceiver at a central command and control location. We emphasize that the remote monitoring and configuration (parameters like data rates, frequency, among others.) capability enabled by the GUI will ease the operator load by alleviating the need to physically travel to the deployed locations. The functioning and design of the GUI has been described in detail in Section \ref{Sec:GUI}.

We further reemphasize that the software-defined stack is designed to be reconfigurable to modify the cross-layer routing as the requirements evolve in the future or depending on the desired network application. It is noteworthy that the utility function discussed above is representative of one such example of the algorithm where the network application desires energy-aware routing. This specific choice was application-specific.\textit{ The broad impact of this work is the reconfigurable nature of the software-defined stack such that it keeps evolving to meet the future requirements rapidly.}

\subsection{Graphical User Interface Application}\label{Sec:GUI}

\begin{figure*}
    \centering
    \includegraphics[width=2 \columnwidth]{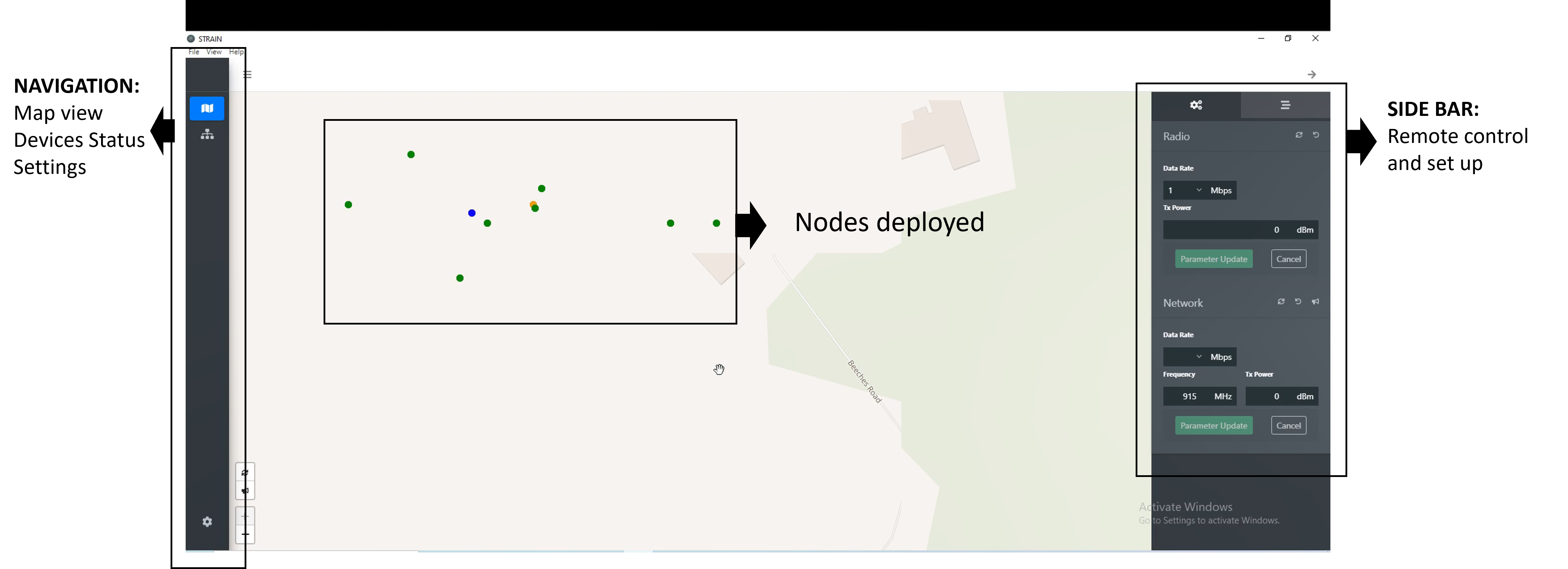}
    \caption{Graphical User Interface (GUI) Application}
    \label{fig:GUI}
\end{figure*}

In this article, we provide an example of how an application such as a GUI can be configured for the network using the proposed software architecture. As described in the previous section, remote network (re-)configuration and monitoring are achieved with the GUI shown in Figure \ref{fig:GUI} coupled with a central node (referred to as gateway node). The gateway node is the only node that directly communicates with the GUI (which runs on a host PC) via Ethernet. The GUI issues command packets to the gateway node which are then transmitted to the other nodes over the air. Multiple packet types are designed to issue specific commands. These packets are divided by their purpose and target (i.e. entire network or just a particular node). Essentially, each of the new packet types is associated with a command and can have a unicast or broadcast variant depending on the intended destination(s) of the packet. 

The GUI uses a build system of Babel and Webpack. This ensures the GUI has a small footprint and is both faster to develop on and faster to run as an end-user. The GUI is split into two processes: \textit{Main} and \textit{Renderer}. The \textit{Main} process handles the communication to the network (through the gateway node). The \textit{Renderer} process handles user input and displaying output information that is used to interact with the GUI. When the user wants to issue a command to the network, they are required to enter the value they would like to change and click the button labeled \textit{Update Parameter}. Once the \textit{Update Parameter} button is clicked, the \textit{Renderer} process of the GUI will create a control packet and send it to the \textit{Main} process of the GUI. Upon receiving the control packet, the \textit{Main} process will send it to the gateway node on a ZMQ Dealer/Dealer socket. The GUI is constantly listening to the Dealer socket for any packet sent from the network.

After receiving the control packet, the gateway node will extract the command and target fields of the packet and use them to determine what type of packet - broadcast or unicast - needs to be sent. For example, the command could be \textit{Get Status} and the target could be all the nodes in the network which will cause the gateway node to respond by issuing the \textit{Get Status} command for all nodes using a broadcast packet.

A broadcast packet is broadcasted for all nodes in range of the gateway node and then rebroadcast by each node that receives it for the first time. Upon receiving a broadcast packet, a node will process the packet, strip its header, and send the packet payload up to user-space for packet handling. A unicast packet is sent to a target node on the route determined by the SEEK routing daemon using the target IP field from the GUI's command packet. 
The reset command is utilized with two different packet types, \textit{Reset Node} and \textit{Reset Network}. The \textit{Reset Network} packet is a broadcast packet and is used to force each node on the network to restart the protocol stack. Restarting involving safe closing and exiting of all active daemons and starting the software stack again. The \textit{Reset Node} packet is a unicast packet that forces a specific target node to restart.

There are two packets that utilize the \textit{Get Status} command, Request Network Status and Request Node Status. A Request Network Status packet is a broadcast packet that is sent upon the gateway receiving a \textit{Get Status} command from the GUI that targets all nodes on the STRAIN network. The Request Node Status packet is a unicast packet that is sent upon the gateway node receiving a \textit{Get Status} command for a target node. Upon receiving one of these packets from the gateway, the receiving node will send a \emph{Response to Gateway} packet as an acknowledgment to the gateway. 

A \textit{Response to Gateway} packet will include OAI about the receiving node including the node's IP address and GPS location. Once the responses reach the gateway node, they are sent back to the GUI via the dealer/dealer ZMQ socket. The GUI will use this information to display the node's status. If after a set duration, a node's response does not reach the GUI, the status of the node will be set to \textit{Disconnected} in the GUI. An example of the list of nodes in the network displaying its status can be seen below in Figure \ref{fig:DeviceStatus}. The battery bar being grey, indicates DC power in addition to battery power. No battery bar indicates only DC power, and a green bar is only battery power. The list of devices also includes action buttons that are \textit{Get Network Status}, \textit{Reset Network}, and \textit{Announce Gateway} respectively. These are displayed in the upper right of the table.

\begin{figure}
    \centering
    \includegraphics[width=1 \columnwidth]{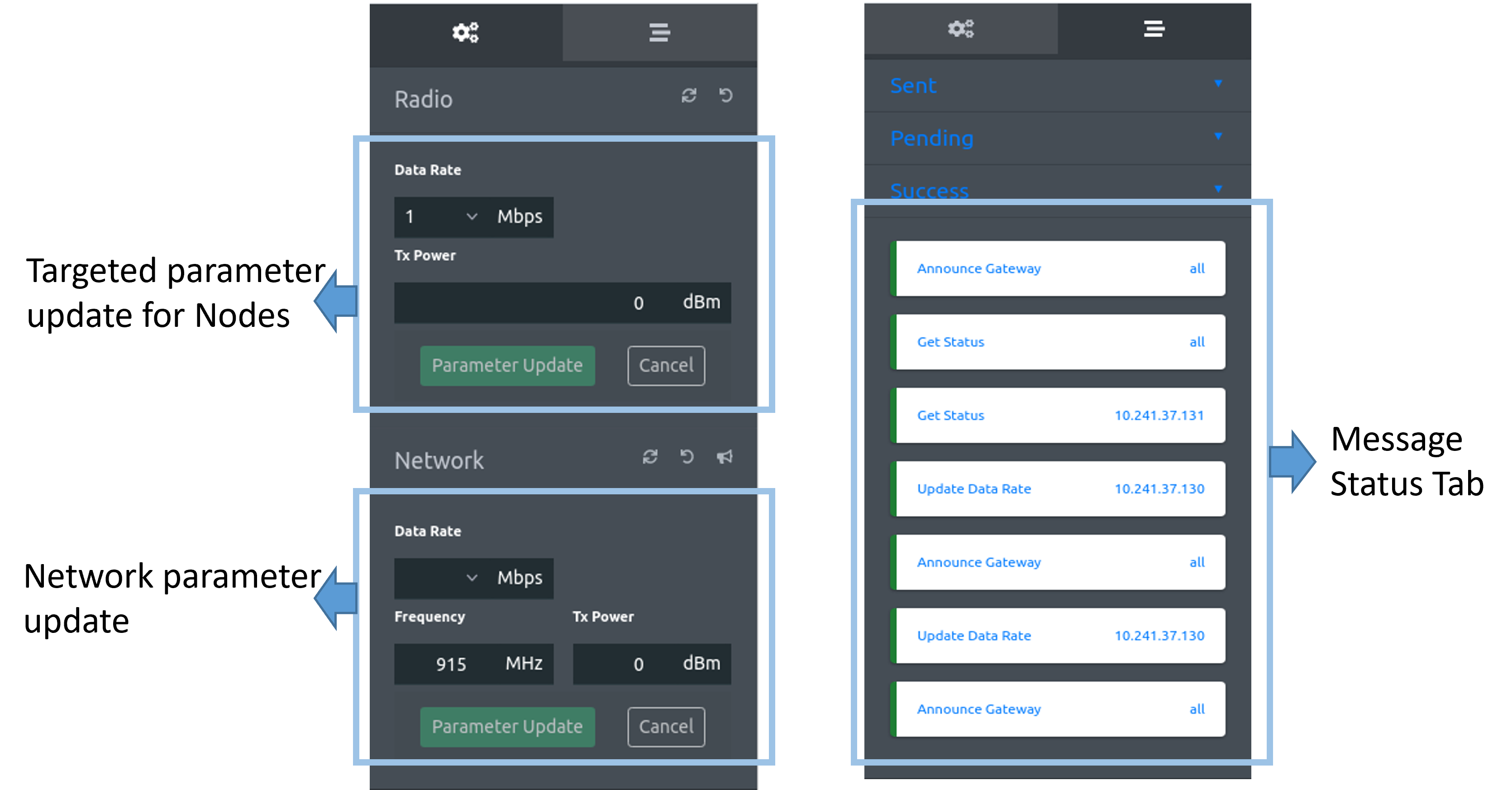}
    \caption{Sidebar of the GUI}
    \label{fig:GUI_SideBar}
\end{figure}

Figure \ref{fig:GUI_SideBar} shows the \textit{Message List} tab of the sidebar with the \textit{Success} section toggled. It is used to view the status of packets sent from the GUI. The \textit{Sent} section displays packets sent from the GUI. The \textit{Pending} section displays packets that are yet to reach the gateway node. The \textit{Success} section is where packets that have received a response appear. \textit{All} indicates the packet sent was a broadcast packet intended for every node in the network. An\textit{ Announce Gateway} for \textit{All} will be listed under success when the gateway successfully receives a response from any node in the network. An IP address, such as "10.241.37.131``, indicates the packet was unicast. A \textit{Get Status} for "10.241.37.131`` will be listed under \textit{Success} once an acknowledgement from the node at IP address "10.241.37.131`` is received at the Gateway. Figure \ref{fig:GUI_SideBar} also shows the \textit{Config} tab of the sidebar in the GUI and is used to execute commands on a selected node. Figure \ref{fig:GUI} shows the map view in the GUI which reflects the network topology. The gateway node is displayed as a blue dot to make it evident to the user.

\begin{figure}
    \centering
    \includegraphics[width=1 \columnwidth]{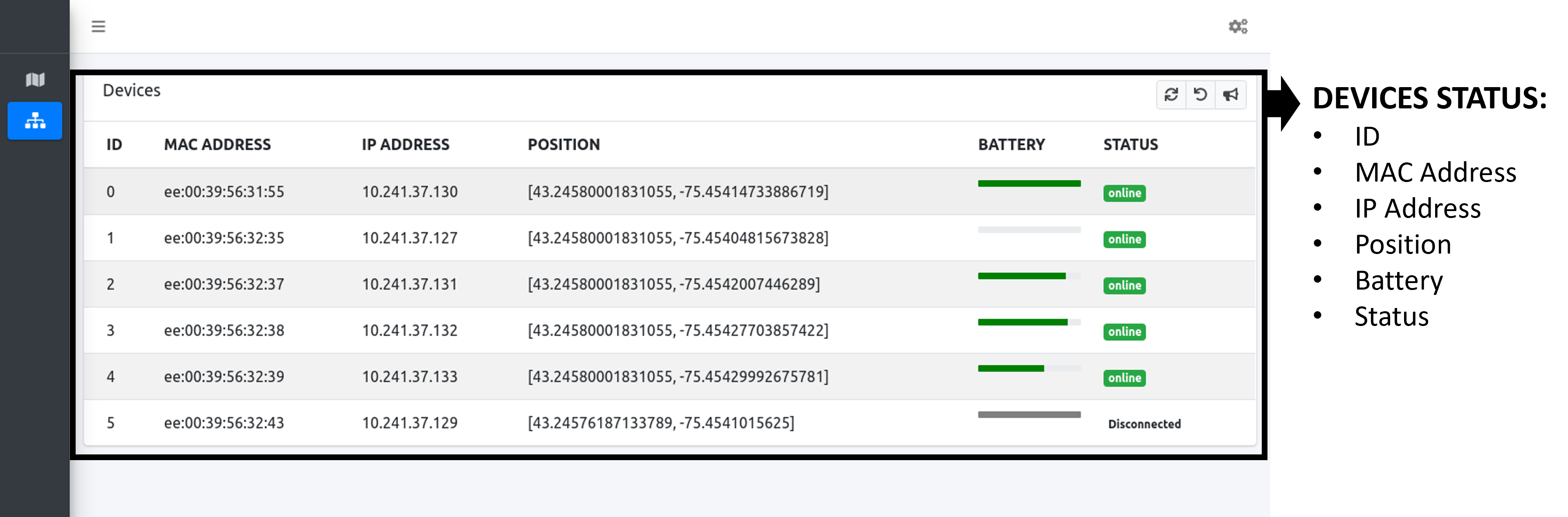}
    \caption{GUI's device Status Page}
    \label{fig:DeviceStatus}
\end{figure}

Since the gateway node is the only node that can directly communicate with the GUI, the gateway must recognize every new node added to the network. Upon startup, each new node broadcasts a \textit{New Node Announcement} packet to neighboring nodes requesting the gateway node IP. This request is broadcast until a neighboring node responds with the requested IP in a \textit{New Node Announcement Response} packet. As a result, the new node and the neighboring nodes update the node list and neighbor lists. Once the new node has the gateway IP, it repeatedly sends a \textit{Hello Gateway} packet until the new node is acknowledged by the gateway. Once the gateway receives this \textit{Hello Gateway} packet, the gateway informs the GUI of the new node and sends a \textit{Gateway New Node Acknowledgement} completing the addition of a new node to the network.

A \textit{Set Network Parameter} packet is broadcast to all nodes in range of the gateway node. The three commands currently available with \textit{Set Network Parameter} are \textit{Update Data Rate}, \textit{Update Tx Power}, and \textit{Update Frequency} commands which are sent throughout the network starting from the gateway and being rebroadcast from the nodes until every node has received the command. The command \textit{Update Data Rate} has four values associated with that can be sent through the network. It changes current data rate of every node in the network to the supported values. In this work, it was set to one of the four values, which are \textit{1 Mbps}, \textit{2 Mbps}, \textit{5.5 Mbps}, and \textit{11 Mbps}. The command \textit{Update Tx Power} sets the attenuation for every node in the network based on the frontend supported values. Similarly, the \textit{Update Frequency} command will set the frequency of the nodes to the specified value.
A Set Node Parameter packet is unicast from the gateway node to a targeted node in the network. The commands currently available with Set Node Parameter are \textit{Update Data Rate} and \textit{Update Tx Power}. The \textit{Update Data Rate} and the \textit{Update Tx Power} command have the same possible values as above for Set Network Parameter.

An \textit{Acknowledge Gateway} packet is transmitted by non-gateway nodes after the nodes receive a \textit{Set Network Parameter} or a \textit{Set Node Parameter} packet. The \textit{Acknowledge Gateway} packet is a unicast packet from the non-gateway node to the destination (gateway) node. Its purpose is to acknowledge the gateway node after making the requested change with \textit{Set Network Parameter} or \textit{Set Node Parameter} packet. It is only created and sent after the change occurs. Summary of remote control and monitoring capabilities are provided in Table \ref{tab:GUI_Command}

\begin{table*}[h!]
 \centering
 \captionof{table}{Summary of GUI Commands}
    \begin{tcolorbox}[tab2,tabularx={|p{5 cm}||p{14 cm}}]
      \textbf{Application Command}  &\textbf{Description}  \\ \hline\hline
   Get Node Status    &Gets the status of the selected node  \\ \hline
   Get Network Status  &Gets the status of the entire networks   \\ \hline
   Announce Gateway  &Announces the gateway node's information to all nodes connected on the network\\ \hline
   Set Data Rate: Apply to Node  &Sets the data rate of the selected node. \\ \hline
   Set Frequency  &Sets the frequency of the network\\ \hline
   Set Transmit Attenuation  & Sets the transmit power of the a target node or the entire network  \\ \hline
   Reset Node  & Forces a selected node to restart   \\ \hline
   Resent Node &  Forces the entire network to restart  \\ \hline
    \end{tcolorbox}{}
    \label{tab:GUI_Command}
\end{table*}




\section{Outdoor Experiments and Results}
\label{Sec:Resutls}

As part of the feasibility analysis, we performed extensive outdoor experiments to evaluate the performance of the implemented software-defined protocol stack in a realistic environment. Since the proposed solution is implemented on an e-SDR, it can be extended to any frequency of interest supported by the target hardware. To accomplish the experiments, we received a temporary experimental license from Federal Communications Commission (FCC) to utilize the 430 MHz frequency within a geographical area. The novelty and utility of the SEEK algorithm itself have been established in previous work \cite{Jagannath19ADH_HELPER, HELPERPatent}. Hence here the focus is to compare various settings and the impact on overall performance (throughput and reliability) of implementing cross-layer optimized algorithms on GPP supported with FPGA-based PHY layer for outdoor deployment. To the best of our knowledge, this is the first time such a comprehensive evaluation has been undertaken using deployment-ready cross-layer optimized e-SDRs. The default parameters are listed in the Table \ref{tab:para} unless otherwise specified.

\begin{table}[h!]
    \centering
    \caption{Parameters Of Outdoor evaluation\label{tab:para}}
    \begin{tcolorbox}[tab2,tabularx={|p{3 cm}||p{5 cm}}]
      \textbf{Parameters}   & \textbf{Values}  \\ \hline\hline
      Data rates   & 1, 2, 5.5, 11 Mbps   \\ \hline
    Transmit Power & 1-3.5 W (based on attenuation)   \\ \hline
    Center Frequency & 430 MHz \\ \hline
    Bandwidth & 22 MHz \\ \hline
    Payload Size   & 1000 Bytes   \\ \hline
    Segment Size   & 32 Packets   \\ \hline
    \end{tcolorbox}{}
\end{table}

\subsection{Transmission Range Evaluation}

\begin{figure}
    \centering
\includegraphics[width=.99 \columnwidth]{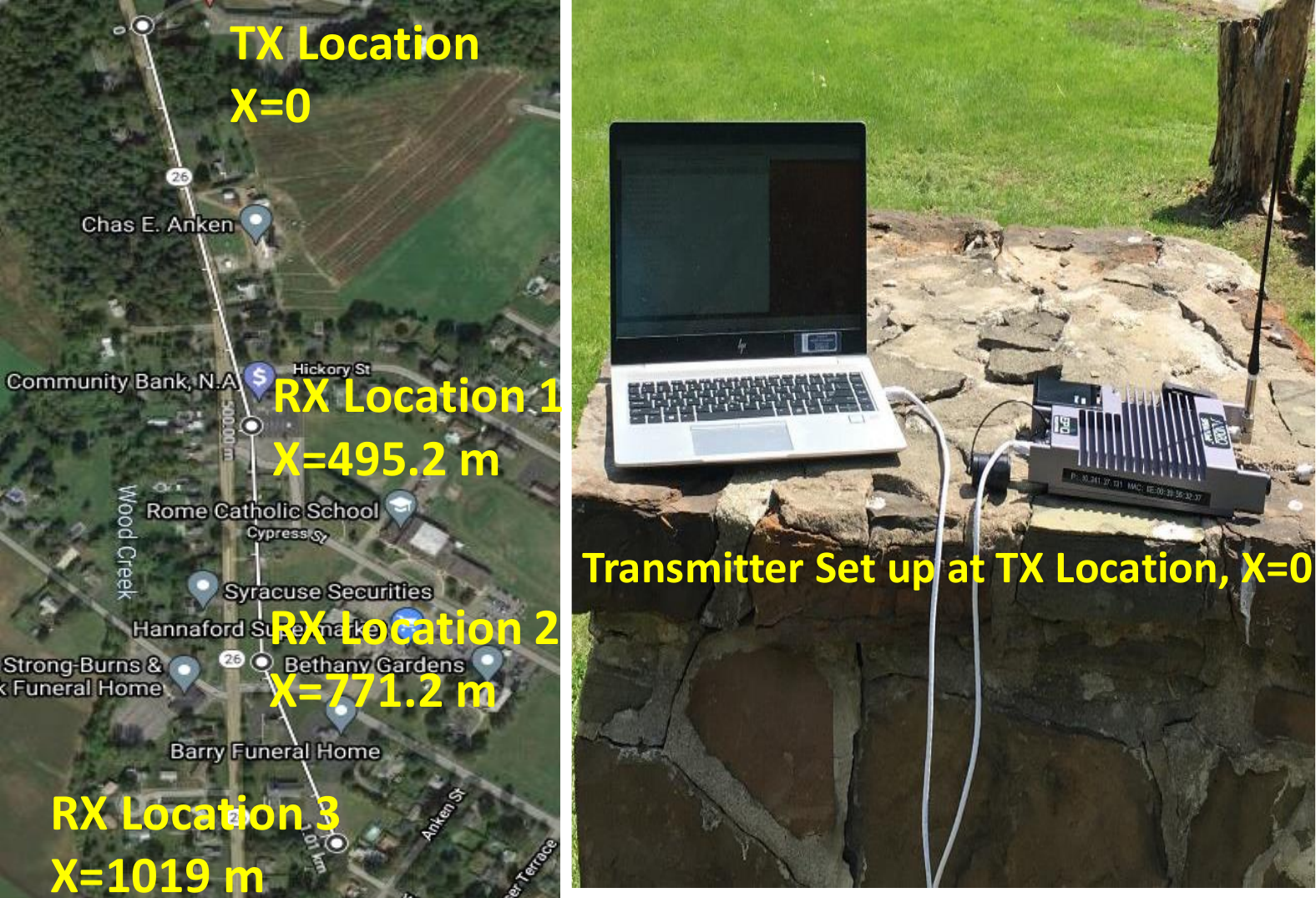} 
\caption{Range testing setup and map-view of the locations.}
\label{fig:RangeTesting}
\end{figure}

The target deployment scenario for the device is remote test and evaluation sites, hence, the design goal for transmission range was to achieve up to $1$ km. The setup of the transmitter and the key testing locations from our range testing experiments is shown in Figure \ref{fig:RangeTesting}. Since the authorization to transmit was only for a limited geographical area, we redesigned the MAC protocol to ensure the receiving node does not transmit any control packet and just acts as a receiver for the purpose of range testing. This ensured that the receiver that needs to be outside the geographical locations does not transmit and just acts as a passive receiver to evaluate the reliability of the link from the transmitter. The rest of the software protocol stack was the same as the architecture discussed in Section \ref{Sec:SystemDesign}. For each reported value in the Table \ref{tab:Range_Test}, the results were averaged over 10,000 packets transmitted in each run. \textit{We define \textbf{reliability} as the percentage of packets received with respect to packets sent.}

\begin{table}[h!]
    \centering
    \caption{Link reliability for varying transmission ranges}
    \label{tab:Range_Test}
    \begin{tcolorbox}[tab2,tabularx={|p{2 cm}||p{2.7 cm}||p{2.5 cm}}]
      \textbf{Data Rate}   & \textbf{Distance} &\textbf{Reliability}    \\ \hline \hline
    \multirow{4}{*}{2 Mbps}  & 495.2 m  &   99.9\%   \\ \cline{2-3}
      & 771.2 m  &  99.77\%   \\ \cline{2-3}
      & 1019 m  &   98.08\%   \\ \cline{2-3}
\hline
    \multirow{4}{*}{5.5 Mbps}  & 495.2 m  &  99.62\%   \\ \cline{2-3}
      & 771.2 m  &  96.03\%   \\ \cline{2-3}
      & 1019 m  &  97.16\%   \\ \cline{2-3}
\hline
      \multirow{4}{*}{11 Mbps}  & 495.2 m  &  85.28\%   \\ \cline{2-3}
      & 771.2 m  &  31.56\%   \\ \cline{2-3}
      & 1019 m  &  13.9\%   \\ \hline
    \end{tcolorbox}{}
\end{table}

Table \ref{tab:Range_Test} shows that the performance was consistently good for both $2$ Mbps and $5.5$ Mbps up to $1019$ m but the performance of $11$ Mbps degraded sooner. Due to the FCC license restrictions and urban environment, the next feasible point was at $1942$ m from the transmitter at which point no packets were received. The performance degradation of $11$ Mbps was attributed to sensitivity to lower signal-to-noise-ratio and multipath propagation effects. This was further substantiated since the performance did improve when we tested a smaller payload size. For example, reducing the payload size to 100 Bytes at a distance of $1019$ m increased the reliability from $13.9\%$ to $30\%$. Further, we also analyzed the Received Signal Strength Indicator (RSSI) values which provided us insight that the performance of the physical layer was deteriorating at RSSI of $\sim15$ dB or higher than that observed during wired in-laboratory experiments. This finding revealed there is room for improvement in sensitivity ($\sim15$ dB) in future iterations of the FPGA PHY. Overall, the experimentation was successful as we hit our target distance of 1 km even at $5.5$ Mbps. 

\subsection{Peer-to-Peer Experiments}

\begin{figure*}[h!]
\begin{minipage}[h]{0.2 \linewidth}
\centering
\includegraphics[width=.95 \columnwidth]{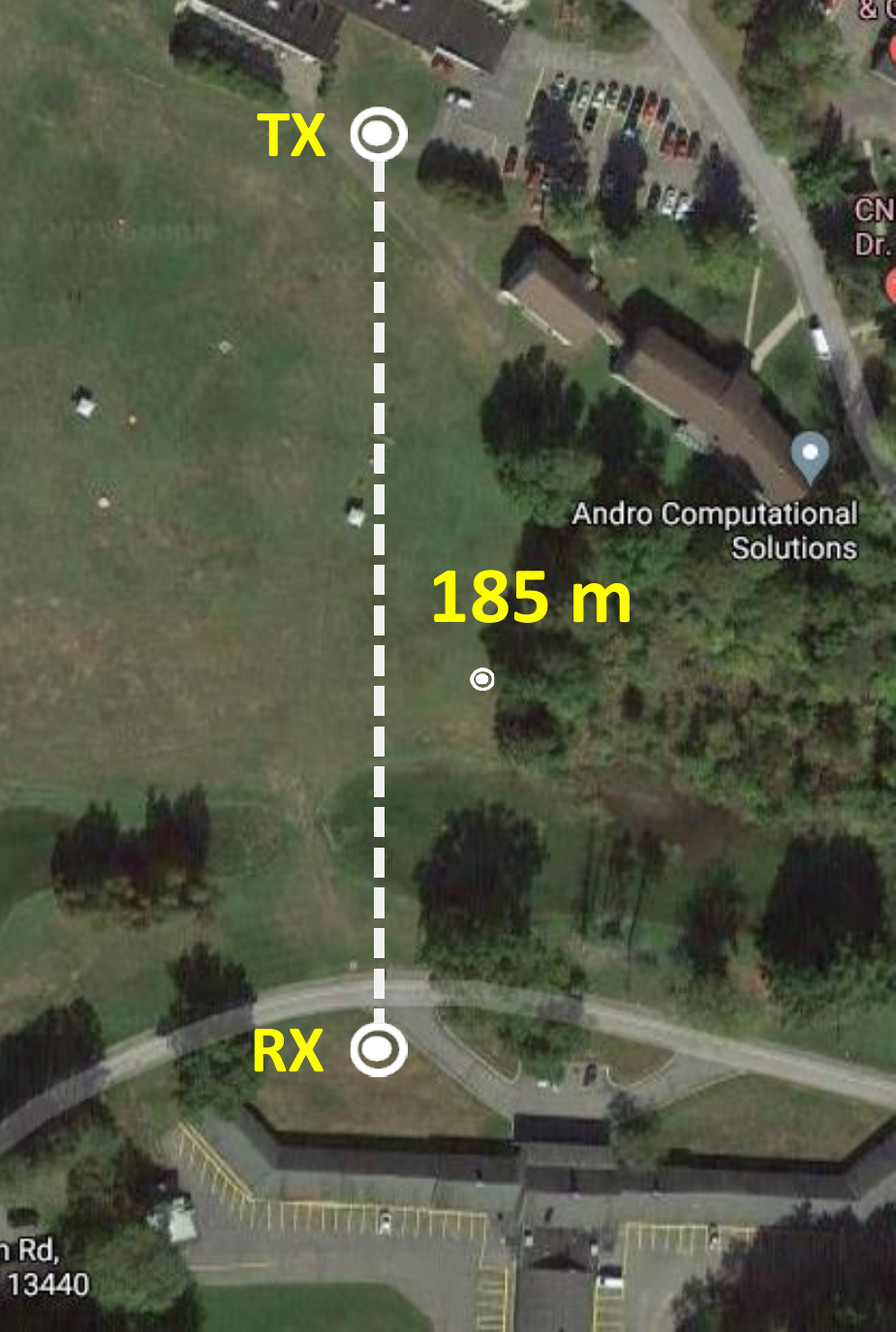} 
\caption{Peer-to-Peer Topology}
\label{fig:P2PMap}
\end{minipage}
\begin{minipage}[h]{0.39 \linewidth}
\centering
\includegraphics[width=.95 \columnwidth]{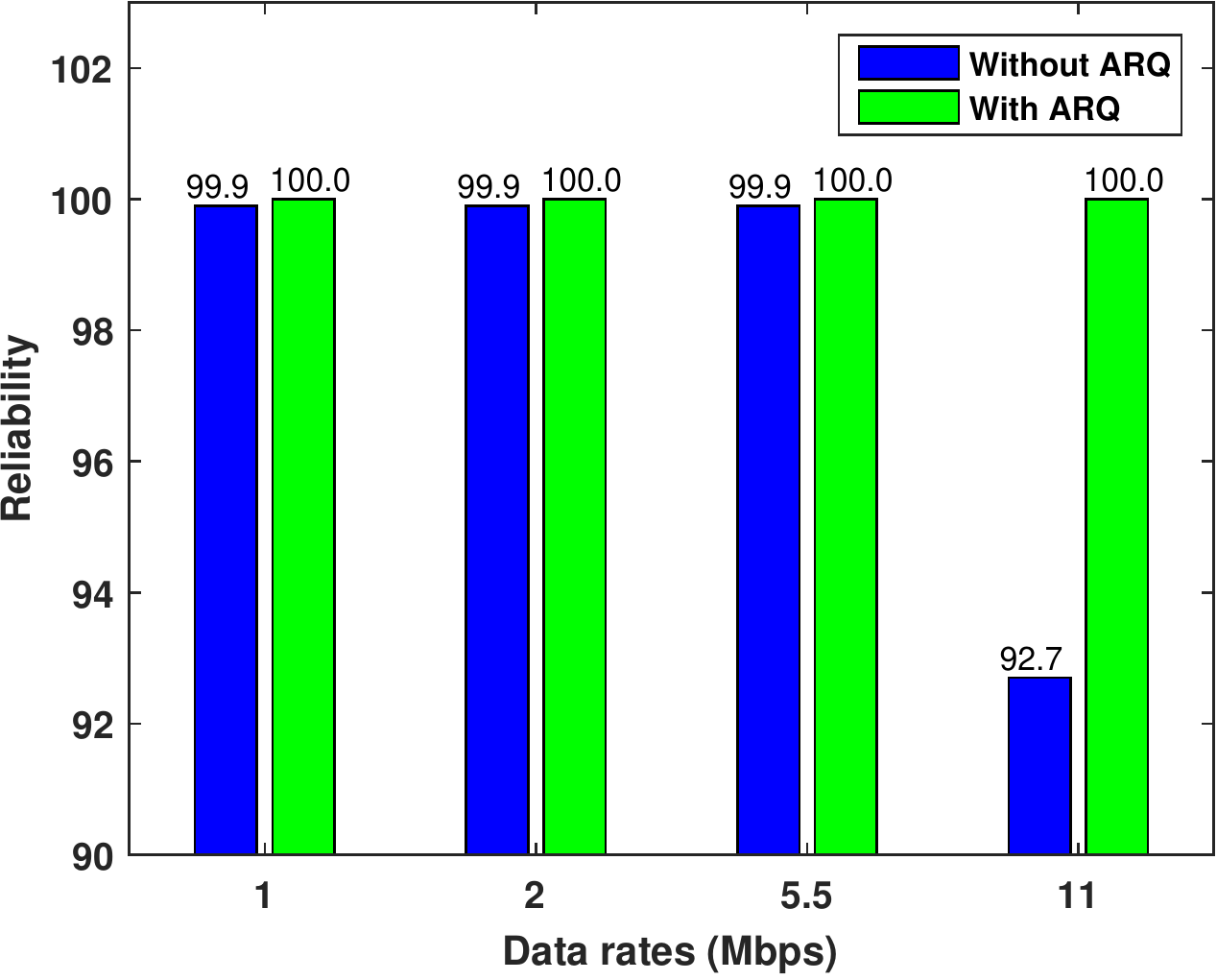} 
\caption{Reliability vs data rates}
\label{fig:RL_peer}
\end{minipage}
\begin{minipage}[h]{0.39 \linewidth}
\centering
\includegraphics[width=.95 \columnwidth]{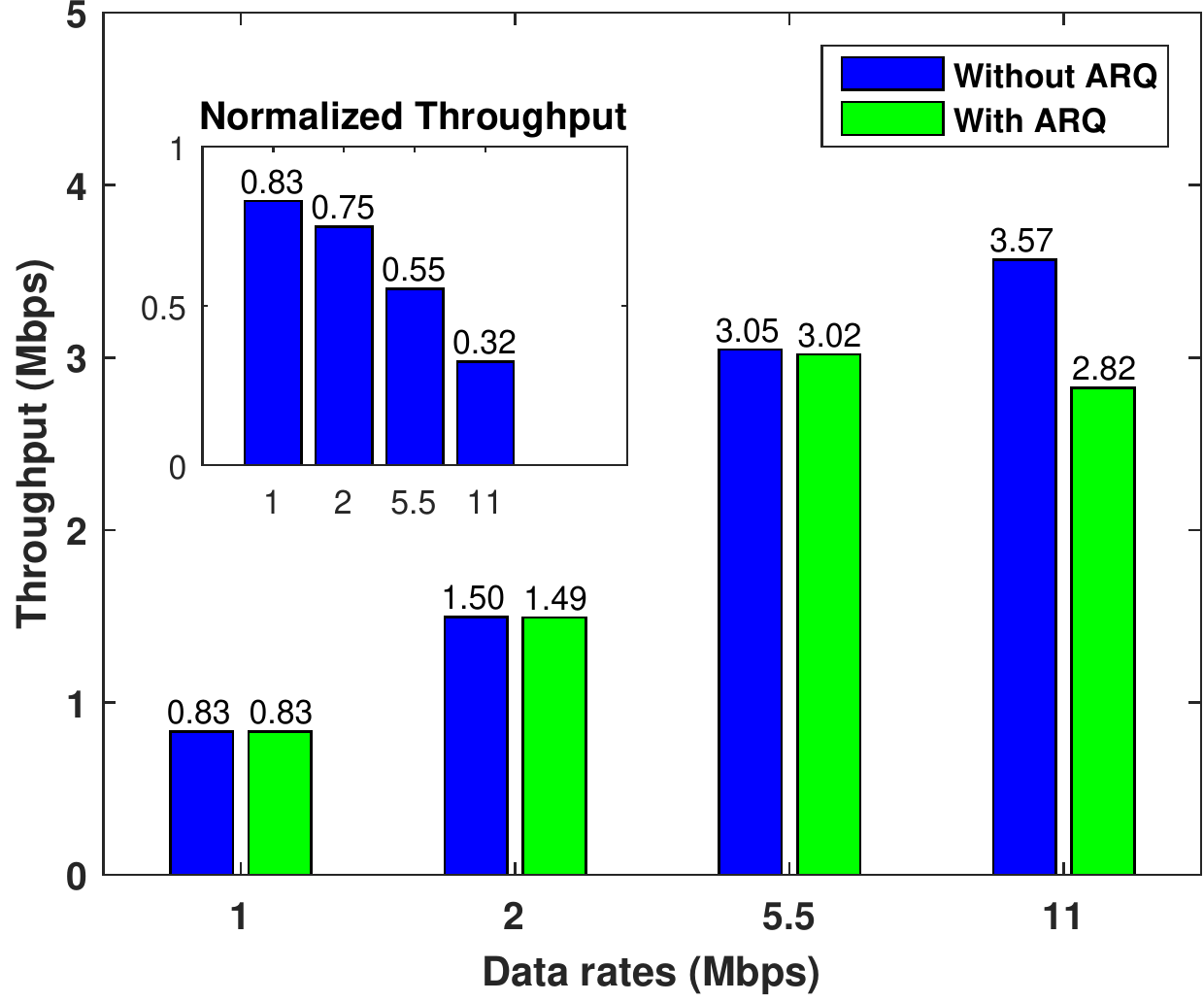} 
\caption{Throughput vs data rates}
\label{fig:Th_peer}
\end{minipage}
\end{figure*}

\begin{figure}
    \centering
\includegraphics[width=.9 \columnwidth]{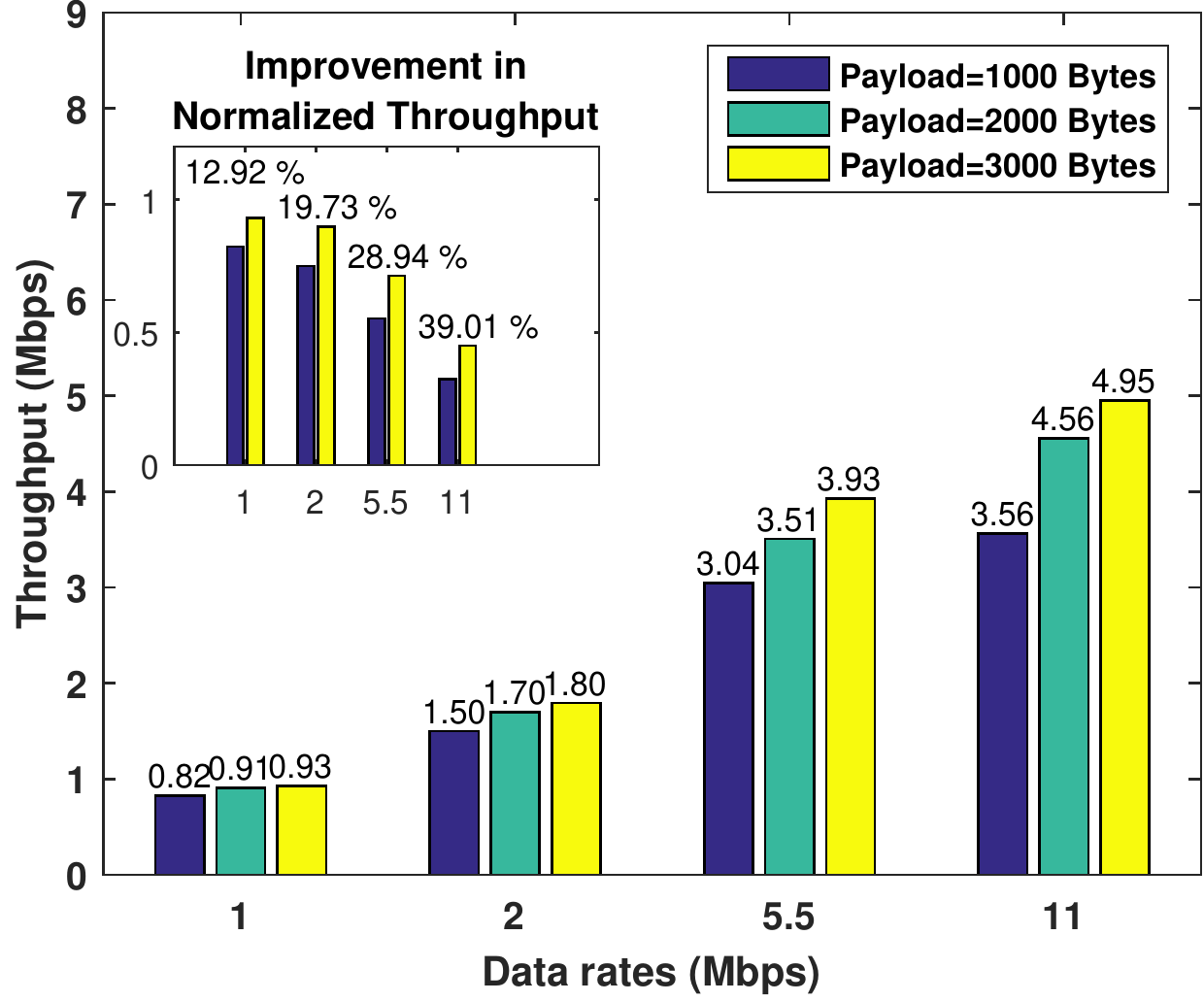} 
\caption{Throughput for varying payload}
\label{fig:Th_payload}
\end{figure}

As described earlier, the decision to implement the PHY on FPGA and the rest of the protocol stack was to achieve the best combination of high data rates and the required flexibility to implement cross-layer routing protocols. Yet, it is important to study the delays and associated overhead it takes to make the necessary calls between the GPP and FPGA to support packet transmission and how it impacts the data rates. To accomplish this, we perform outdoor peer-to-peer testing for varying data rates and payload sizes to analyze the impact. The outdoor locations of the transmitter and receiver are depicted in Figure \ref{fig:P2PMap}.

First, in Figure \ref{fig:RL_peer}, we study the reliability at the different data rates with and without automatic repeat request (ARQ) enabled at the MAC layer. It can be clearly seen that the transceiver has high reliability close to $100\%$ for 1, 2, and 5.5 Mbps even without ARQ enabled. At 11 Mbps the reliability decreases to $92\%$ without ARQ and returns to $100\%$ with ARQ by trading off throughput. Figure \ref{fig:Th_peer} depicts how the peer-to-peer throughput varies at different PHY data rates 1, 2, 5.5, 11 Mbps. \textit{The \textbf{throughput} in this case only considered the payload of the packets and does not include the headers or the control packets}. The throughput considered here is also referred to as goodput in some literature. The figure also depicts the \textit{\textbf{normalized throughput} which is defined as the ratio between throughput and data rate}. The design choice provides solid performance at 1 and 2 Mbps especially considering it depicts the goodput achieved. The impact of the overhead becomes visible as the data rate increases to 5.5 and 11 Mbps. This hypothesis is further substantiated in our experiments that show an increase in normalized throughput with the increase in the payload as shown in Figure \ref{fig:Th_payload}. The normalized throughput increased by up to $39\%$ for 11 Mbps when the payload was increased from $1000$ Bytes to $3000$ Bytes.

\subsection{6-Node Line Network}

\begin{figure}
    \centering
    \includegraphics[width=1 \columnwidth]{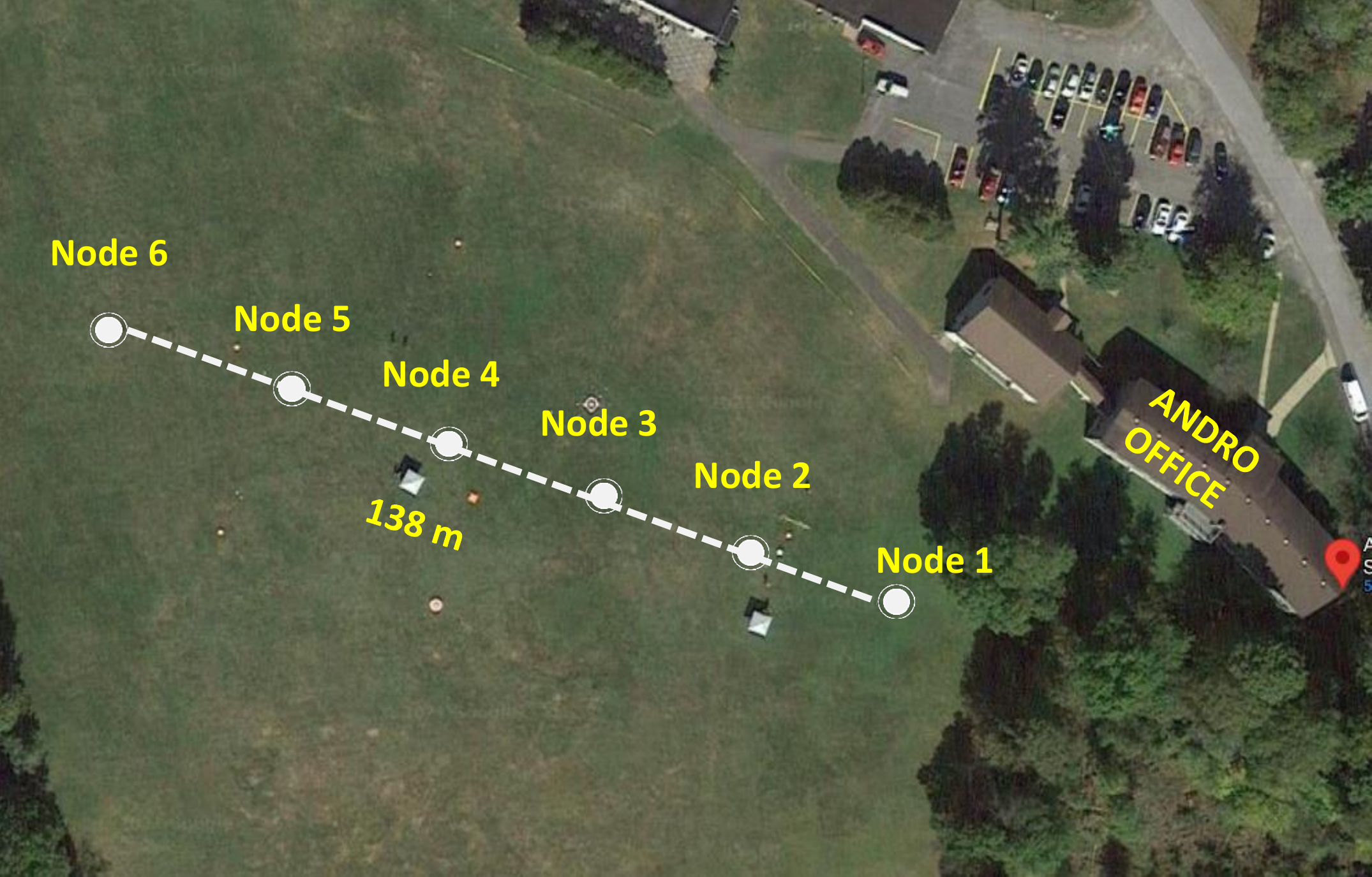} 
    \caption{6-node Line Network Topology}
    \label{fig:LineMap}
\end{figure}

\begin{figure*}[h!]
\begin{minipage}[h]{0.49 \linewidth}
\centering
\includegraphics[width=.9 \columnwidth]{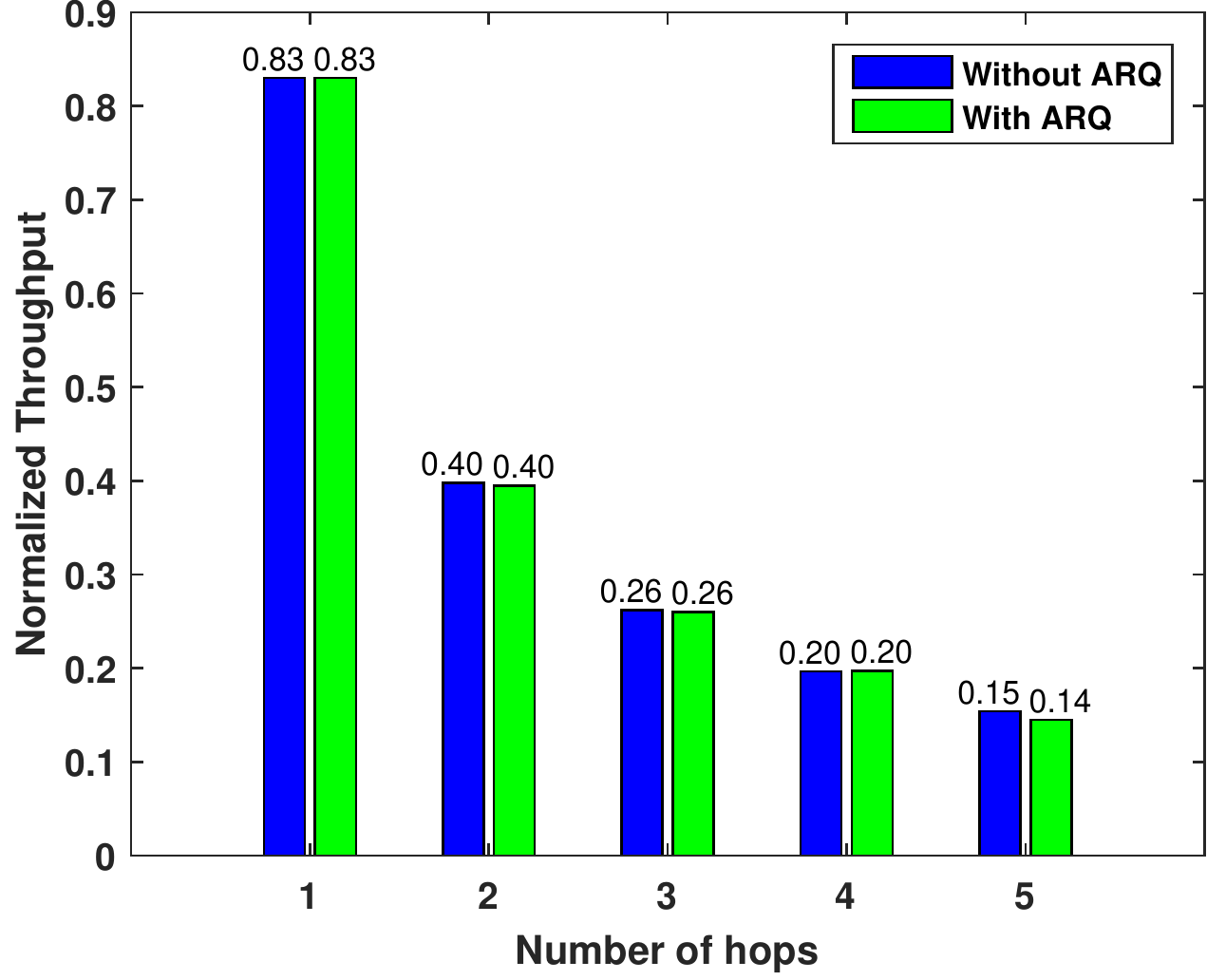} 
\caption{Line Network Throughput Analysis}
\label{fig:Line_TH}
\end{minipage}
\hspace{0.1 cm}
\begin{minipage}[h]{0.5 \linewidth}
\centering
\centering
\includegraphics[width=.9 \columnwidth]{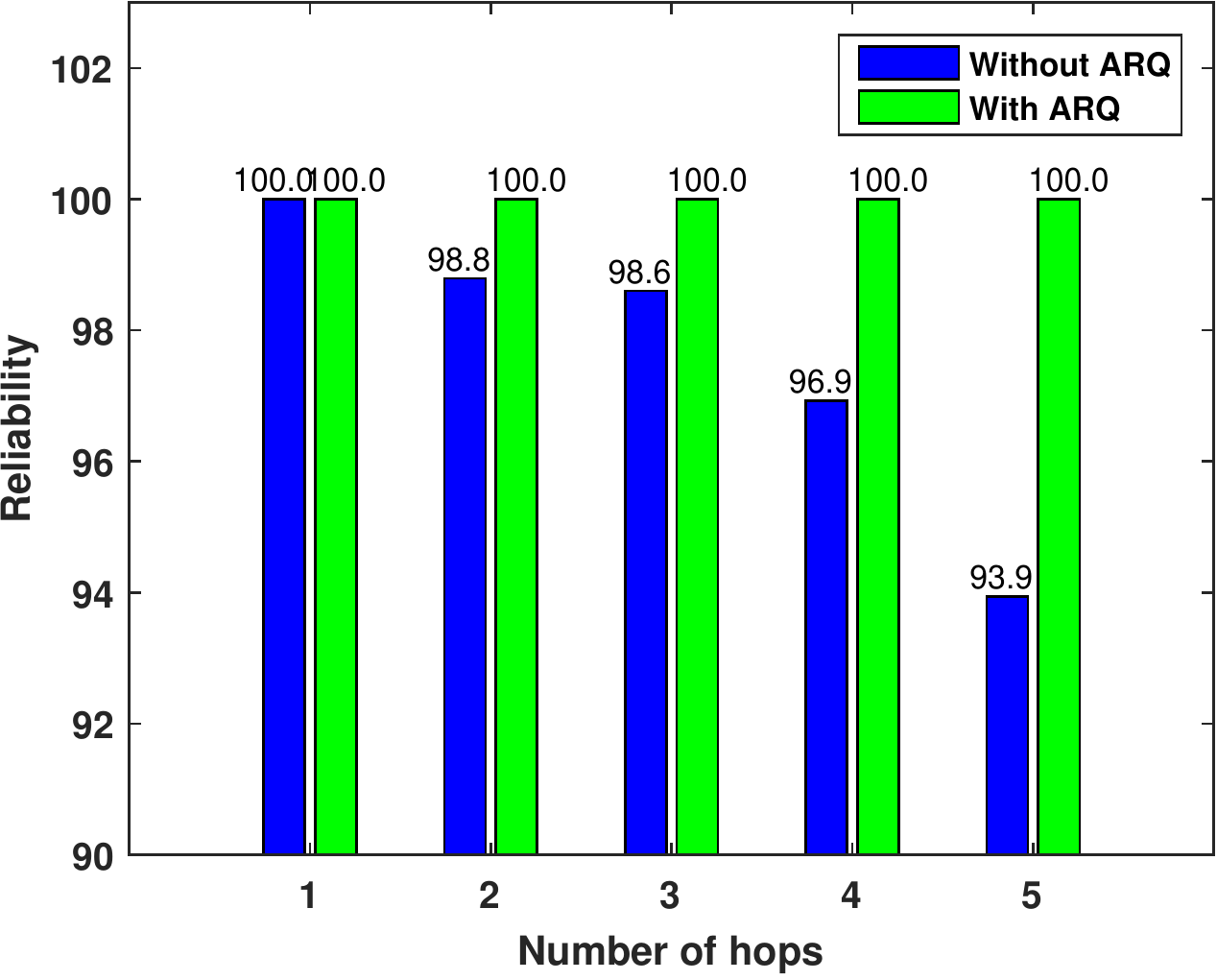} 
\caption{Line Network Reliability Analysis}
\label{fig:Line_RL}
\end{minipage}
\end{figure*}

In the next set of experiments, we examined the impact of the number of hops on normalized throughput and the reliability of the network. The line network topology is shown in Figure \ref{fig:LineMap}. In this topology only the nodes adjacent to each other are neighbors. Hence five hops are required from node 1 to node 6. The experiments were conducted with and without ARQ. To ensure a line network in the limited geographical area authorized for experiments by the FCC, we had to set neighbor tables in software. This also implied that all nodes will cause interference to each other which also serves as a good test for ad hoc channel access capabilities/performance. The results followed exactly the hypothesized pattern in the case of normalized throughput. In the single-hop a normalized throughput of 0.83 and all the other measures for hops 2, 3, 4, 5 followed close to 1/2, 1/3, 1/4, and 1/5 of the normalized throughput of the single-hop case as shown in Figure \ref{fig:Line_TH}. This implied that the nodes were utilizing (sharing) the spectrum efficiently without too much loss due to collision. As shown in Figure \ref{fig:Line_RL}, the reliability also seemed to confer with this finding as in most cases reliability was over $95\%$. The lowest reliability was still $93.9\%$ for the 5-hop case even when the ARQ was disabled. All the reliability was back to $100\%$ when ARQ was enabled. There was only a marginal drop in normalized throughput when ARQ was enabled which was mainly attributed to the link being very reliable even without ARQ leading to only a few retransmissions even when ARQ was enabled.

\subsection{5-Node Network Experiment - Dynamic Routing}

\begin{figure*}[h!]
\begin{minipage}[h]{0.49 \linewidth}
\centering
\includegraphics[width=1 \columnwidth]{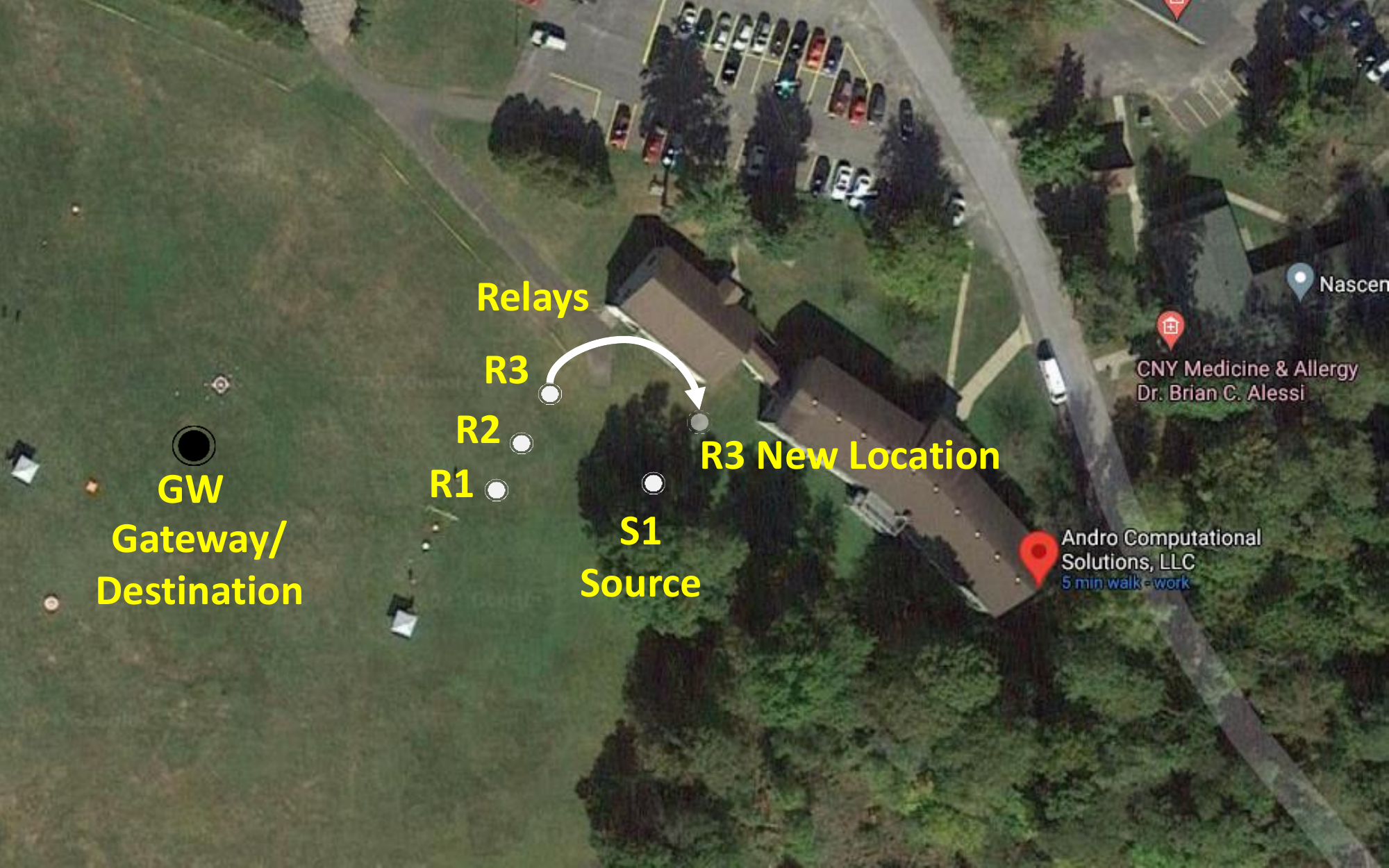} 
\caption{5-node Outdoor Topology}
\label{fig:RouteMaps}
\end{minipage}
\hspace{.1 cm}
\begin{minipage}[h]{0.5 \linewidth}
\centering
\includegraphics[width=1 \columnwidth]{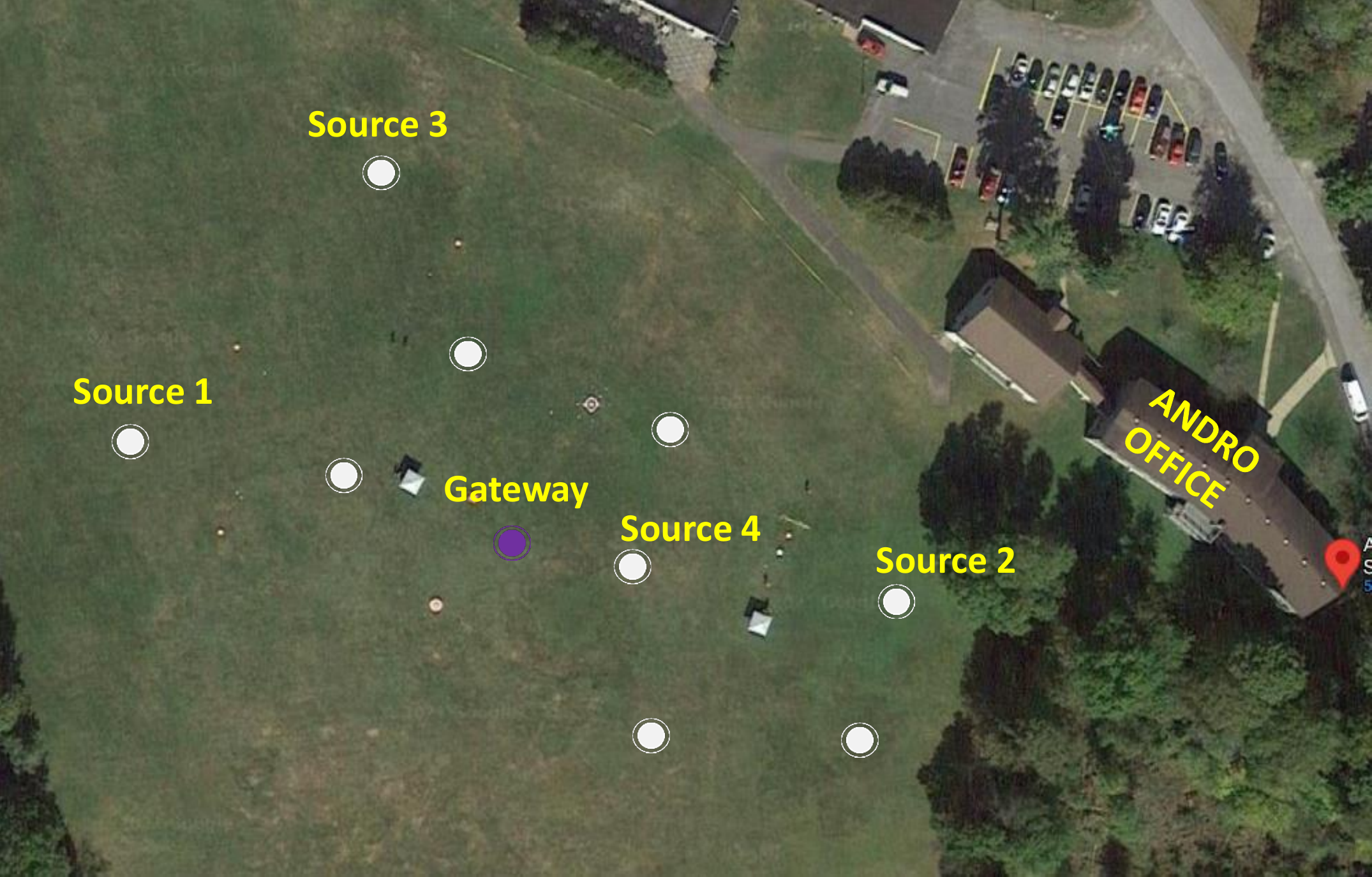} 
\caption{10-node Outdoor Topology}
\label{fig:NetworkMaps}
\end{minipage}
\end{figure*}

For timely evaluation, we designed a specific experiment such that we can determine the effectiveness of the implementation of SEEK algorithm to dynamic changes in the network. To accomplish this, we consider a topology shown in Figure \ref{fig:RouteMaps}. The source \textit{S1} continuously transmits packets destined for the gateway node \textit{GW}. \textit{S1} chooses the appropriate relay among \textit{R1}, \textit{R2}, and \textit{R3} based on SEEK algorithm. We plot the average packets received from each relay node every $10$ s interval. The values are averaged using a moving window of $60$ s duration to get smoother curves.

\begin{figure*}[h!]
\begin{minipage}[h]{0.55 \linewidth}
\centering
\includegraphics[width=1 \columnwidth]{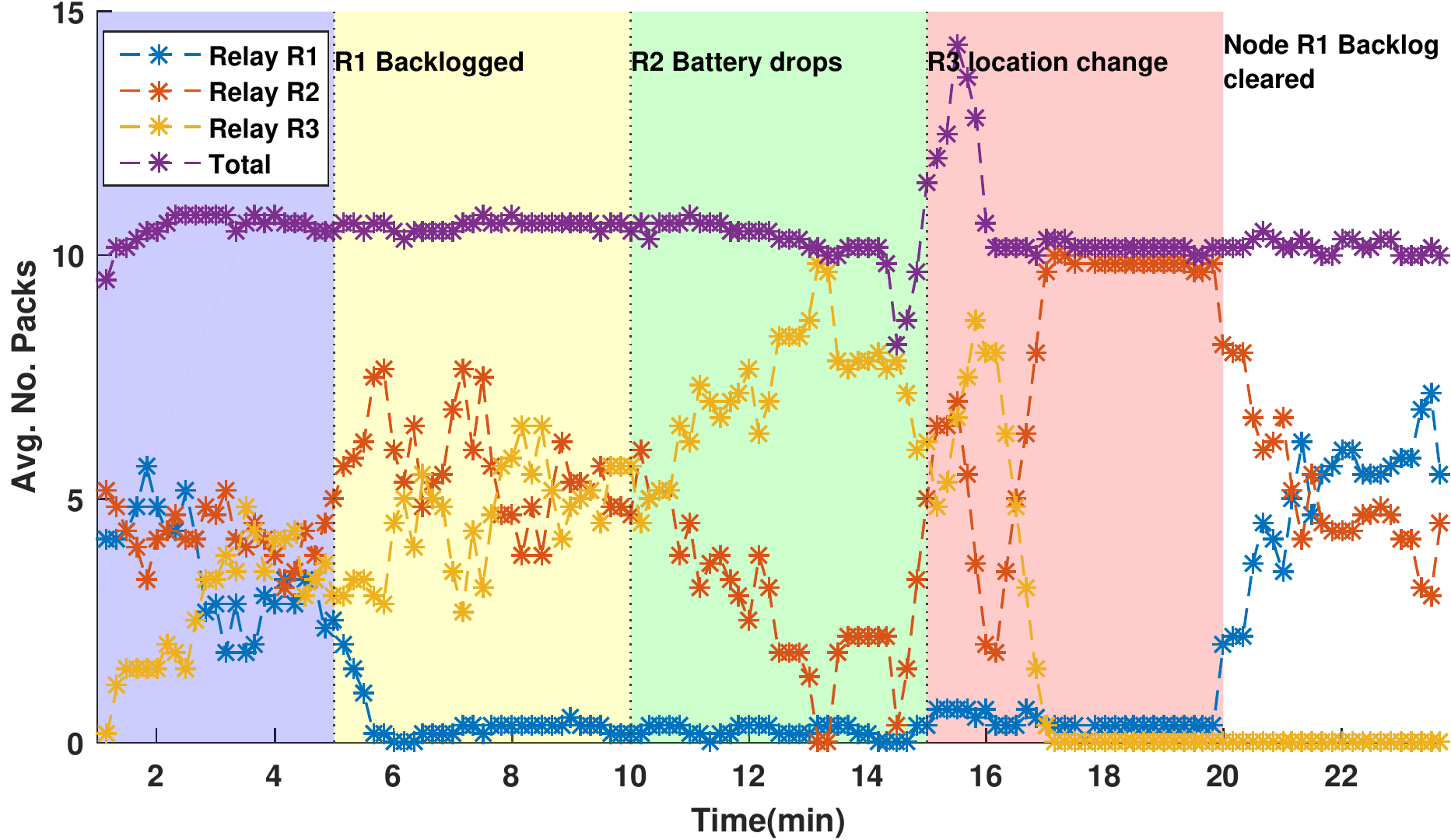} 
\caption{5-node Experiment (dynamic routing)}
\label{fig:route}
\end{minipage}
\hspace{0.5 cm}
\begin{minipage}[h]{0.4 \linewidth}
\centering
\centering
\includegraphics[width=.99 \columnwidth]{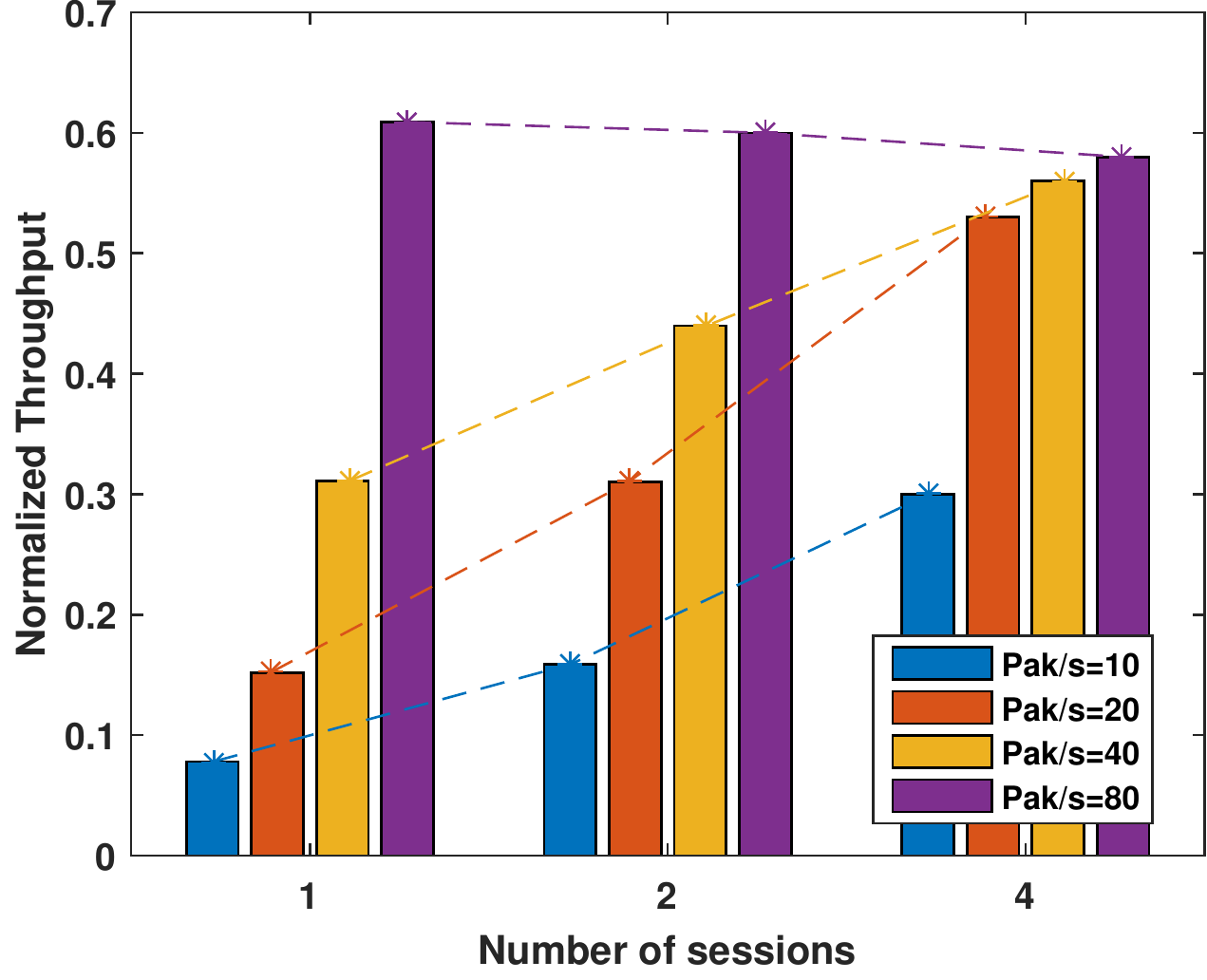} 
\caption{10-node Experiment}
\label{fig:Network}
\end{minipage}
\end{figure*}

In the first part of Figure \ref{fig:route} (blue shade) depicts a scenario where the relay nodes (i.e. nodes with possible forward progress for \textit{S1}) have approximately uniform parameters. As expected, in this scenario, the traffic is evenly distributed between the three relays. Next, we significantly increase the backlog of the \textit{R1} to emulate a congestion scenario at a relay node. As it can be seen (yellow shade), \textit{S1} learns to avoid \textit{R1} and routes the packets through the other two options, \textit{R2} and \textit{R3}. At the beginning of the experiment, \textit{R2} was set up with a battery with low residual energy but was connected to a DC power supply. When a node is connected to a DC power supply, it is equivalent to having the battery completely charged with full residual energy. Now in the next part of the evaluation, the residual energy of \textit{R2} is rapidly dropped by removing the DC power supply and the residual energy in the battery was close to $10\%$. As it can be seen, \textit{S1} recognizes this in the next part (yellow shade) and prefers \textit{R3} so that lifetime of the network can be extended. We remind the readers that the lifetime of the network is defined as the duration of network operation till the first node in the network is completely depleted of residual energy. Therefore, the nodes use the utility function described earlier in Section \ref{Sec:SystemDesign} trying to jointly consider multiple parameters including residual energy. Therefore, one goal is to distribute traffic over multiple paths so that nodes in the network can share the load based on available energy resources to extend the lifetime of the network. This is the exact intuitive behavior observed here in this region of Figure \ref{fig:route}. Thereafter, in the next part, we now move the \textit{R3} node behind the source \textit{S1} as shown in Figure \ref{fig:RouteMaps} such that it is no more a feasible next-hop that provides forward progress. This information is quickly realized by the source \textit{S1} through the beacon packets as in all the previous cases. In this situation, with \textit{R1} congested, \textit{R3} is no more the possible choice for forward progress, \textit{S1} must return to \textit{R2} with a lower battery since that is the only possible choice to relay the packets to the destination (\textit{GW}). Hence in the red region of Figure \ref{fig:route}, it can be seen that the traffic through \textit{R2} picks up while the traffic through \textit{R3} drops. Another interesting observation is that when the R3 initially changes location it still has some packets remaining for the \textit{GW}, and \textit{GW} is still within its transmission range. Accordingly, after the initial drop in traffic, it still is able to get those packets to the \textit{GW} which can also be seen as a spike in the total traffic momentarily to overcome the dip it experienced a few moments prior. In the final section of the Figure \ref{fig:route}, the backlog of \textit{R1} is cleared. Now, in this case, the \textit{S1} realizes that it does not have to use the \textit{R2} with lower residual energy anymore and can share the traffic load through \textit{R1}. As a result, it can be seen that now traffic through \textit{R1} increases while the traffic through \textit{R2} reduces. \textit{It is also interesting to see that even with all these changes, the total packets/s at the \textit{GW} (represented at the "total" legend in Figure \ref{fig:route}) remains very stable demonstrating the rapid adaptability of the cross-layer optimized nodes.} In this experiment, we have demonstrated the gamut of routing decisions the nodes can make in a distributed manner by just gathering information from their immediate neighbors demonstrating the effectiveness of the cross-layer optimized routing using an intuitively representative experiment.

\subsection{10-Node Network Experiments - Network Capacity}

In this experiment, we evaluate how the network handles the increasing number of sessions and source (packet generation) rate. The 10-node topology is shown in Figure \ref{fig:NetworkMaps}. The experiments were conducted at $1$ Mbps of data rate, the source rate was varied from $[10, 20, 40, 80]$ packets/s, and the number of sessions (independent sources generating traffic to the gateway node) was set to $1$, $2$ and $4$. This is motivated by a typical use case for our intended deployment scenario. As expected, Figure \ref{fig:Network} demonstrates that the normalized throughput increases and gets saturated both when the source rate increases and the number of sessions increase. There is no drop in performance even when there are multiple sessions ($4$) with high source rates ($80$) and the normalized throughput is maintained at the saturated value ($\sim0.6$). This shows the efficiency of the network in handling multiple traffics and session rates as expected even when the traffic increases. 

\section{Conclusion and Future Direction}\label{Sec:Conclusion}

This article introduces the first-known fieldable cross-layer optimized, embedded software-defined radio (e-SDR) that has been designed, developed, and matured to provide high throughput and reliability by implementing the protocol stack (except the physical layer) on an embedded ARM processor. To emphasize the contribution of the article, we survey the literature to highlight that the majority of cross-layer optimized techniques have been either limited to simulations or extended to hardware testbeds. We point out some of the key hurdles that might have led to the halt in the maturity of these solutions. The survey also demonstrates the large scale of applicable domains for cross-layer optimization if the solutions are designed and matured to be deployed in real-life applications. 

First, to overcome the hardware-related challenges, we discuss how a COTS e-SDR can be customized and re-configured for deploying cross-layer protocol stack. We then provide a detailed discussion on the software architecture and design choices that ensured the feasibility of such an endeavor. Ensuring the right balance between modular and flexible architecture along with efficiency is the key takeaway for this discussion. Most importantly, we have demonstrated through extensive outdoor field experiments the capability, range, and dynamic routing under varying network conditions. 

In the future, cross-layer optimized networking can be a key enabler for several applications in IoT, wireless sensor networks, 5G, and beyond. For example, device-to-device communication can be enhanced by cross-layer optimization that could be implemented on resource constrained edge platfroms. Additionally, several of the tactical applications rely on ad hoc networking to provide the much needed flexibility during tactical operations. We hope this successful demonstration of cross-layer optimized e-SDR will provide directions for future cross-layer optimized solutions to benefit tactical and commercial applications. 

\section*{Acknowledgment and Disclaimer}

The authors would like to thank John Orlando, Jeff Porter of Epiq Solutions for their support, Raymond Shaw of Spectrum Bullpen for help with Spectrum Supportability Risk Assessments, and Dan O' Connor of ANDRO Computational Solutions for his help during the outdoor testing.

(a) Contractor acknowledges the Government's support in the publication of this paper. This material is based upon work supported by the US Army Contract No. W15P7T-20-C-0006. (b) Any opinions, findings, and conclusions or recommendation expressed in this material are those of the author(s) and do not necessarily reflect the views of the US Army.

\bibliographystyle{IEEEtran}	
\bibliography{Andro1}


\begin{IEEEbiography}[{\includegraphics[width=1in,height=1.5in,keepaspectratio]{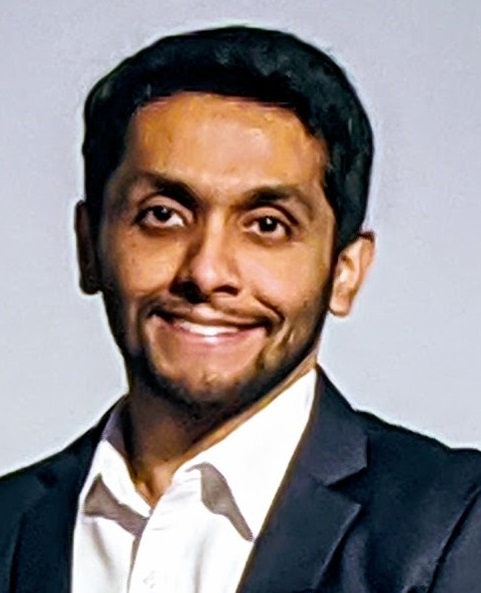}}]{Jithin Jagannath} (SM'19) is the Chief Technology Scientist and Founding Director of the Marconi-Rosenblatt AI/ML Innovation Lab at ANDRO Computational Solutions. He is also the Adjunct Assistant Professor in the Department of Electrical Engineering at the University at Buffalo, State University of New York. Dr. Jagannath received his B. Tech in Electronics and Communication from Kerala University;  M.S. degree in Electrical Engineering from University at Buffalo, The State University of New York; and received his Ph.D. degree in Electrical Engineering from Northeastern University. He is an IEEE Senior member and serves as IEEE Industry DSP Technology Standing Committee member. He also serves on the Federal Communication Commission's (FCC) Communications Security, Reliability, and Interoperability Council (CSRIC VIII) Working Group 1. Dr. Jagannath was the recipient of the 2021 IEEE Region 1 Technological Innovation Award with the citation, "For innovative contributions in machine learning techniques for the wireless domain''. 

Dr. Jagannath heads several of the ANDRO's research and development projects in the field of Beyond 5G, signal processing, RF signal intelligence, cognitive radio, cross-layer ad-hoc networks, Internet-of-Things, AI-enabled wireless, and machine learning. He has been the lead and Principal Investigator (PI) of several multi-million dollar research projects. This includes a Rapid Innovation Fund (RIF) and several Small Business Innovation Research (SBIR)s for several customers including the U.S. Army, U.S Navy, Department of Homeland Security (DHS),  United States Special Operations Command (SOCOM). He is currently leading several teams developing commercial products such as SPEARLink\texttrademark, DEEPSpec\texttrademark~ among others. He is the inventor of 11 U.S. Patents (granted, pending, and provisional). He has been invited to give various talks including Keynote on the topic of machine learning and Beyond 5G wireless communication. He has been invited to serve on the Technical Program Committee for several leading technical conferences.

\end{IEEEbiography}

\begin{IEEEbiography}[{\includegraphics[width=1in,height=1.25in,clip,keepaspectratio]{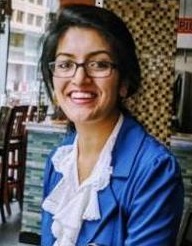}}]{Anu Jagannath} received her MS degree (2013) from State University of New York at Buffalo in Electrical Engineering. She is currently a Senior Scientist at ANDRO Computational Solutions, LLC and serves as the Associate Director of Marconi-Rosenblatt AI/ML Innovation Lab at ANDRO. Her research focuses on MIMO communications, Deep Machine Learning, Reinforcement Learning, Adaptive signal processing, Software Defined Radios, spread spectrum systems, LPI/LPD communications, spectrum sensing, adaptive Physical layer, and cross-layer techniques, medium access control and routing protocols, underwater wireless sensor networks, signal intelligence and so on. She serves as author and coauthor for book chapters and multiple research publications in journals and conference proceedings. She has rendered her reviewing service for conferences such as IEEE Annual Consumer Communications \& Networking Conference (CCNC) and IEEE International Workshop on Signal Processing Advances in Wireless Communications (SPAWC). She is the co-Principal Investigator (co-PI) and Technical Lead in multiple Rapid Innovation Fund (RIF) and SBIR/STTR efforts involving developing embedded MIMO solutions, deep and reinforcement learning for wireless communications, signal intelligence, and mesh networking.
\end{IEEEbiography}

\begin{IEEEbiography}[{\includegraphics[width=1in,height=1.5in,keepaspectratio]{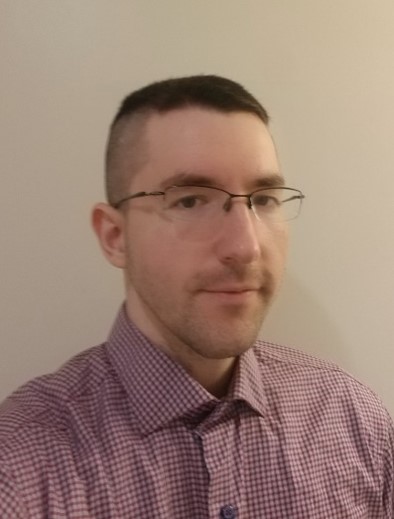}}]{Justin Henney} received his BS from the Rochester Institute of Technology in Electrical Engineering in 2019. He is currently an Associate Scientist/Engineer at ANDRO Computational Solutions, and has worked on developing the SPEARLink\texttrademark~and ARROWLink\texttrademark~products. 
\end{IEEEbiography}

\begin{IEEEbiography}[{\includegraphics[width=1in,height=1.5in,keepaspectratio]{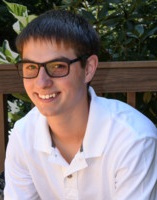}}]{Tyler Gwin} received their B.S. degree in Computer and Information Science from SUNY Polytechnic Institute. He is currently an Associate Scientist/Engineer at ANDRO Computational Solutions, LLC. During his work at ANDRO, he has been part of the team developing SPEARLink\texttrademark~software, along with the research and development of small Unmanned Aerial Systems. Tyler has also assisted in the authoring of a book chapter titled "Deep Learning and Reinforcement Learning for Autonomous Unmanned Aerial Systems: Roadmap for Theory to Deployment." 
\end{IEEEbiography}

\begin{IEEEbiography}[{\includegraphics[width=1in,height=1.5in,keepaspectratio]{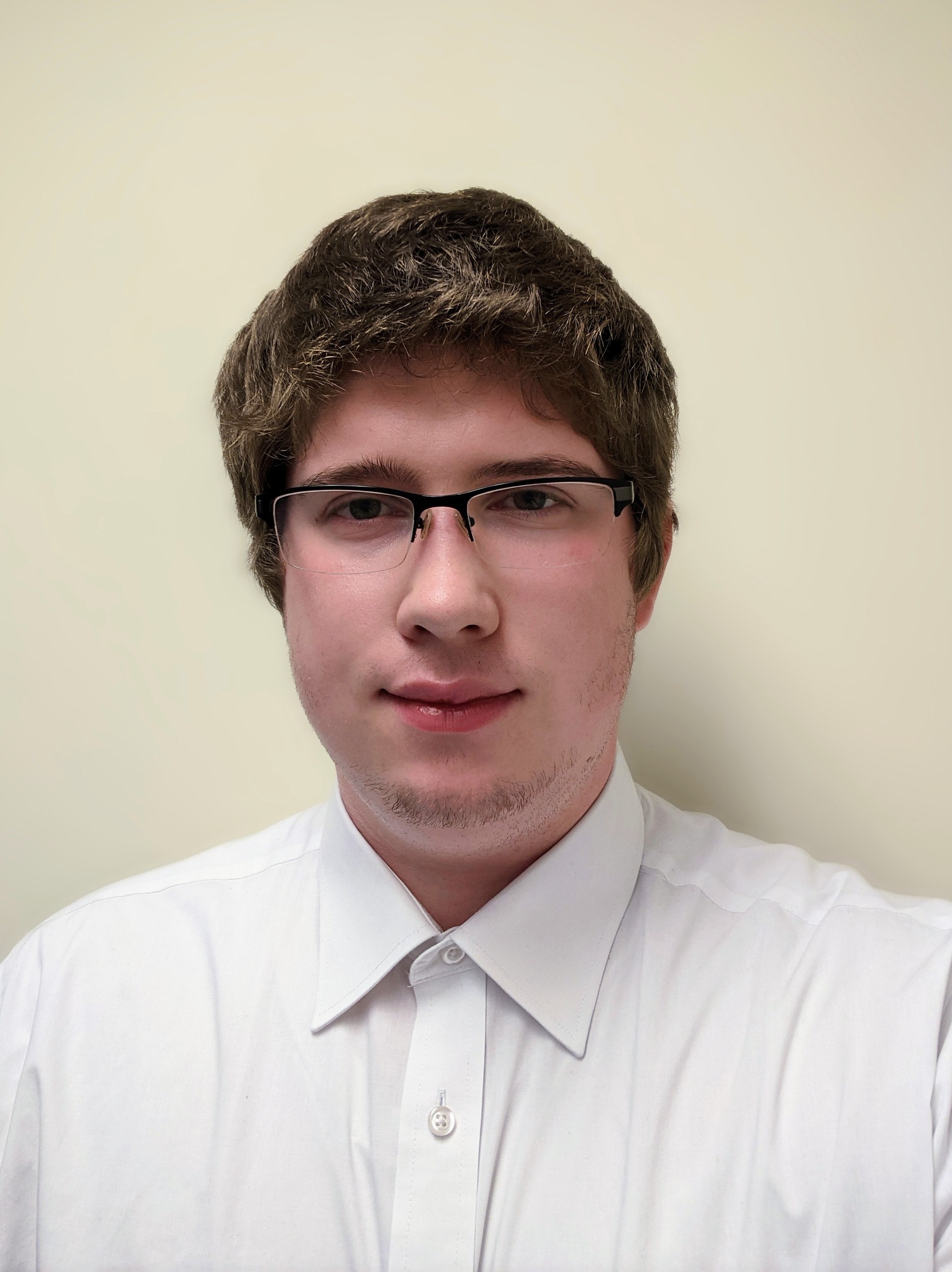}}]{Zackary Kane} received a B.S. in Electrical and Computer Engineer from the State University of New York Polytechnic Institute. Zack is currently a Radio Frequency Engineer/Associate Scientist I at ANDRO Computational Solutions in ANDRO's Marconi-Rosenblatt lab. He has worked on two Small Business Innovation Research (SBIR) projects and the STRAIN Rapid Innovation Fund (RIF) project. While on STRAIN, Zack assisted in range testing the SPEARLink\texttrademark~radio platform and worked on documentation for the SPEARLink\texttrademark~radio. He is currently working on the DeepeRFind Phase I SBIR, generating an database of commercial radio devices, and the I-ROAM RIF, testing and improving the IROAM CDMA physical layer. 
\end{IEEEbiography}

\begin{IEEEbiography}[{\includegraphics[width=1in,height=1.5in,keepaspectratio]{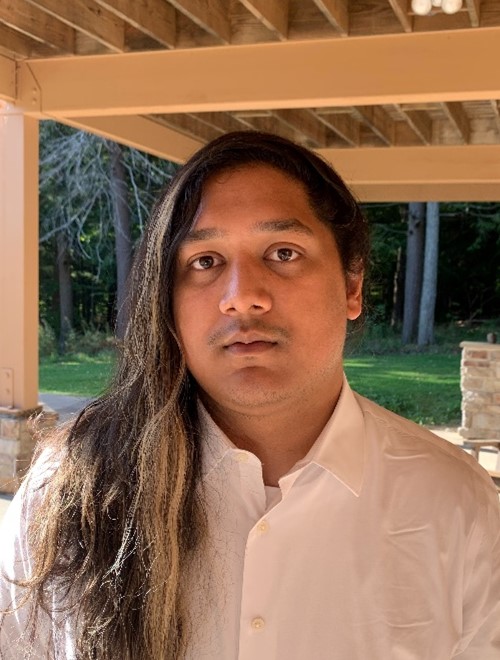}}]{Noor Biswas} received his BS degree in Computer Engineering from the State University of New York Polytechnic Institute in 2020. He is currently an Associate Scientist/Engineer at ANDRO Computational Solutions, and has contributed to projects involving software-defined radio networks. 
\end{IEEEbiography}

\begin{IEEEbiography}[{\includegraphics[width=1in,height=1 in, keepaspectratio]{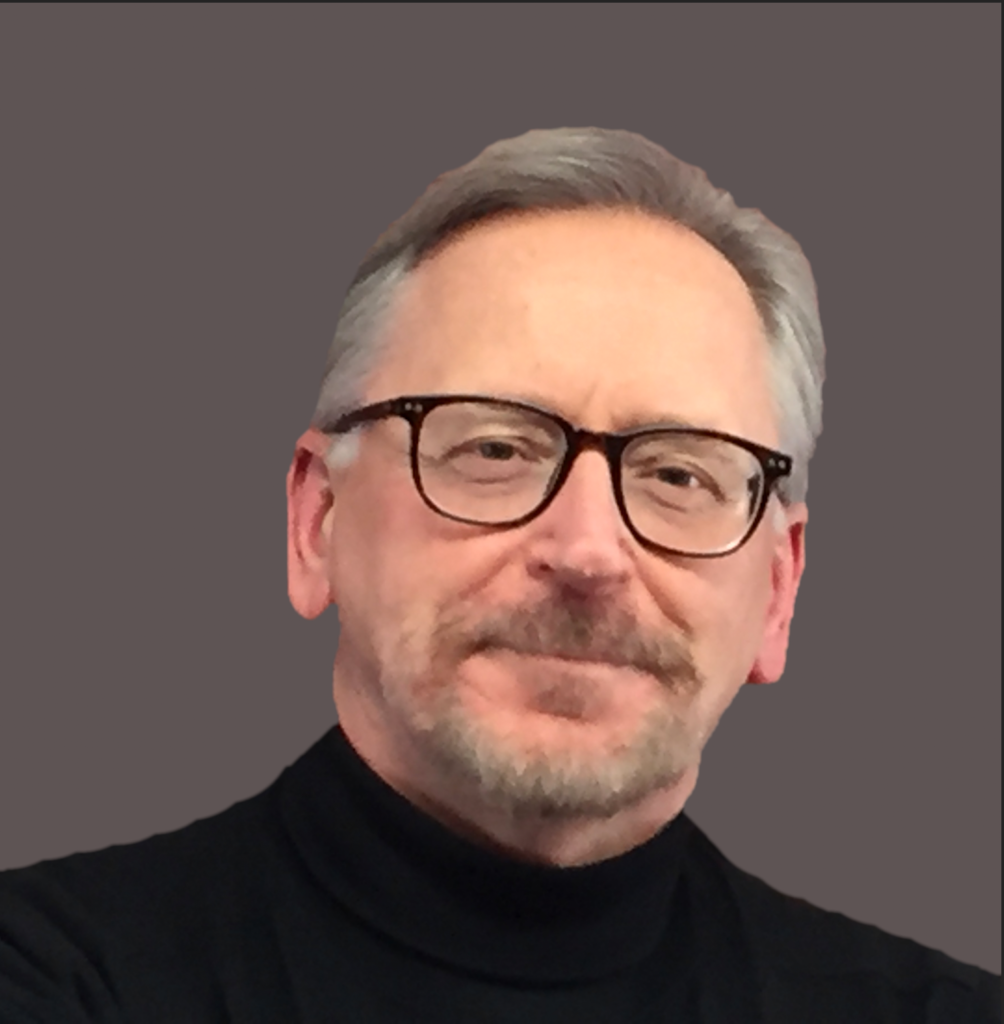}}]{Andrew Drozd} received his Ph.D. from St. John Fisher College in Executive Leadership in Science \& Technology Towards Advancements in Human and Machine Learning Using Theoretic Frameworks for AI/ML Applications. Dr. Drozd has extensive experience in electromagnetics science and technology with expertise in emerging spectrum technologies, radar cross section (RCS), computational electromagnetics (CEM), radar and multisensor data fusion, multitarget identification/tracking, target discrimination, cyber and electronic warfare (EW), and cosite modeling and simulation through his various working relationships over the past 43 years with DoD and Federal government agencies, including the U.S. Army, U.S. Air Force, U.S. Navy, Office of the Secretary of Defense (OSD), Missile Defense Agency (MDA), Defense Information Systems Agency / Department of Spectrum Operations (DISA/DSO), and Department of Homeland Security (DHS).  Over the past several years, Dr. Drozd has led efforts to develop and mature new technologies under Small Business Innovative Research / Small-business Technology Transfer Research (SBIR/STTR) and Commercialization Readiness Program (CRP) contracts. 

Dr. Drozd is an iNARTE (international Association of Radio and Telecommunications Engineers) Certified Engineer (1988-present); IEEE Life Fellow; past president of the IEEE EMC Society (2006-2007); past chairman of the IEEE EMC-S Standards Development and Education Committee (2007-2009); member of the FCC's Communications, Security, Reliability, Interoperability Council (CSRIC VIII - 2021-2023); member of the Alpha Gamma Omicron Chapter of Kappa Delta Pi Honor Society in Education, board member of the Project Fibonacci® Foundation, Inc. focused on STEM leadership education for workforce development; and a 2019 inductee in the Rome Academy of Sciences Hall of Fame.  He holds 10 patents and patents pending including the areas of spectrum governance, signal detection/classification, and block chain auditing. 
\end{IEEEbiography}

\ifCLASSOPTIONcaptionsoff
  \newpage
\fi

\vskip -2.5\baselineskip

\end{document}


\title{\textbf{Cover Letter}}
\maketitle
Due to the 15-count reference limit in the magazine article, all of the related works are not cited in the manuscript. Instead, we have only cited the most relevant ones. This document is provided to serve as a supplementary information to highlight the key contributions of the article and its distinguishing factors with respect to prior research in this field to assist the reviewers and editors. 

There have already been some works on state-of-the-art deep learning approaches as well as future 6G networks. All presented in this document is compared and contrasted here for clarity. 
\begin{enumerate}
    \item The work in [1] studies the key challenges and potential physical layer solutions for 6G. This work talks about algorithmic physical layer approaches and briefly mention deep neural network as a key 6G enabler. However, they do not mention how these physical layer techniques can be realized for an edge computing platform to facilitate the key concept of edge learning in 6G.
    
    \item The authors of [2] present a vision of the 6G network in terms of the various use cases, challenges, and key enabling technologies. However, the focus here is on an overall 6G network and not specifically the physical layer solutions. The authors however mention AI as a pervasive enabler in realizing 6G.
    
    \item In [3], an intelligent edge concept for wireless communication is elaborated. The learning-driven radio resource management and signal encoding problem is studied in depth. However, the study on complexity of the approaches and their implementation platforms (edge computing devices) is lacking. 
    
    \item The authors of [4] presents the various 6G network use cases and enabling technologies. However, the authors only briefly mention integrating AI in the 6G network towards achieving autonomous operation and do not go in-depth into their implementation challenges. 
    
    \item The authors of [5] discusses the 6G architecture and a few promising technologies to realize the 6G requirements. Authors mention the various AI techniques that will serve as key enablers for the intelligent 6G architecture while succinctly mentioning model-driven deep learning approaches as one of the candidate AI techniques.
    
    \item In [6], traditional deep learning for the optical wireless communication system is studied. Specifically, a convolutional autoencoder for image sensor communication is presented and evaluated in simulations. However, the computational complexity and processing delay of such an approach will be very high owing to the maximum likelihood complexity of the decoding layer alone. An in-depth study on the implementation challenges is lacking in this work.
    
    \item The authors of [7] presents the architecture of 6G networks while discussing the concept of intelligent radio. They also briefly mention the importance of an intelligent physical layer in realizing the 6G communication requirements. However, their work does not present the candidate physical layer approaches that will realize the intelligent physical layer concept.
    
    \item In [8], the authors present a detailed study of the traditional machine learning techniques that have been applied to solving wireless communication problems. However, their complexity, latency, and reliability from a hardware-efficient implementation standpoint are lacking in this work.
\end{enumerate}

 Indeed, irrespective of the AI-based upper layer schemes, the communication capabilities in terms of reliability, processing delay, latency, and complexity of a radio is bottlenecked by its underlying physical layer. In this work, we focus on how deep unfolded approaches that integrates deep learning with principled algorithms advances the communication capabilities in realizing a hardware-efficient intelligent physical layer. We aim to motivate the reader in understanding how computational intensive deep learning approaches may not be the solution to achieve the 6G communication requirements. Instead, integrating the powerful learning capability of deep learning with algorithmic knowledge will relieve the computational burden as well as improve the robustness. Therefore, this is the first work that studies the potential intelligent hardware-efficient physical layer solutions that would serve as key enablers to realize the 6G AI radio concept.
\\
\\
\textbf{Related works}

[1] P. Yang, Y. Xiao, M. Xiao and S. Li, "6G Wireless Communications: Vision and Potential Techniques," in IEEE Network, vol. 33, no. 4, pp. 70-75, July/August 2019.
\\

[2] F. Tariq, M. R. A. Khandaker, K. Wong, M. A. Imran, M. Bennis and M. Debbah, "A Speculative Study on 6G," in IEEE Wireless Communications, vol. 27, no. 4, pp. 118-125, August 2020.
\\

[3] G. Zhu, D. Liu, Y. Du, C. You, J. Zhang and K. Huang, "Toward an Intelligent Edge: Wireless Communication Meets Machine Learning," in IEEE Communications Magazine, vol. 58, no. 1, pp. 19-25, January 2020.
\\

[4] M. Giordani, M. Polese, M. Mezzavilla, S. Rangan and M. Zorzi, "Toward 6G Networks: Use Cases and Technologies," in IEEE Communications Magazine, vol. 58, no. 3, pp. 55-61, March 2020.
\\

[5] Y. Zhao, J. Zhao, W. Zhai, S. Sun, D. Niyato, and K-Y. Lam, “A survey of 6g wireless communications: Emerging technologies,” ArXiv, vol. abs/2004.08549, 2020.
\\

[6] H. Lee, S. H. Lee, T. Q. S. Quek and I. Lee, "Deep Learning Framework for Wireless Systems: Applications to Optical Wireless Communications," in IEEE Communications Magazine, vol. 57, no. 3, pp. 35-41, March 2019.
\\


[7] K. B. Letaief, W. Chen, Y. Shi, J. Zhang and Y. A. Zhang, "The Roadmap to 6G: AI Empowered Wireless Networks," in IEEE Communications Magazine, vol. 57, no. 8, pp. 84-90, August 2019.
\\

[8] J. Wang, C. Jiang, H. Zhang, Y. Ren, K. Chen, and L. Hanzo, “Thirty years of machine learning: The road to pareto-optimal next-generation wireless networks,” ArXiv, vol. abs/1902.01946, 2019.


\begin{appendices}
\section{Formulation of optimization problem}\label{Ap:opt}

The global objective of the optimization problem is to find the optimal global vectors $\mathbf{R}$, $\mathbf{F}$ and $\mathbf{P}$ that maximizes the sum of the network utilities, under the constraints of power and $BER$. The formulation of the optimization problem is as follows,

\begin{align}
\mathcal{P}_1\!:\textup{Given}\!&: \mathcal{G(U,E)},\;\;P^{Bgt},\;\; Q_i^s,\;\; BER_{\mathcal{SU}},\;\;BER_{\mathcal{PU}}\notag \\
				\textup{Find}\!&: \mathbf{R},\;\; \mathbf{F},\;\; \mathbf{P},\;\; \mathbf{A}\notag \\
				\textup{Maximize}\!&: \sum_{i \in \mathcal{SU}} \sum_{j \in \mathcal{SU}/i} U_{ij}(a_i^s(t))\\
				\textup{subject\; to}\!&:\notag \\
				& \sum_{s \in S} r_{ij}^s \leq C_{ij}, \forall i \in \mathcal{SU},\; \forall j \in \mathcal{SU} \label{constr:capacity_P1} \\ 
				& SINR_k \geq SINR_{\mathcal{PU}}^{th}(BER_{\mathcal{PU}}), \forall k \in \mathcal{PU}, \forall f \label{constr:PU_P1} \\ 
				& SINR_l \geq SINR_{\mathcal{SU}}^{th}(BER_{\mathcal{SU}}), \forall l \in \mathcal{SU}, \forall f \label{constr:SU_P1} \\ 
				& \sum_{f \in[f_i, f_{i+\Delta f_i}]} P_i(f) \leq P_i^{Bgt}, \forall i \in \mathcal{SU} \label{constr:Power_P1}
\end{align}

In the above formulation, the objective is to maximize the network utility of all the active links. The constraint (\ref{constr:capacity_P1}) restricts the total amount of traffic in link $(i,j)$ to be lower than or equal to the physical link capacity. Constraint (\ref{constr:PU_P1}) and (\ref{constr:SU_P1}) imposes that any transmission by secondary user should guarantee the required $BER$ for the active primary users and secondary user respectively. Finally, $P_i^{Bgt}$ is the instantaneous power available at the cognitive radio.
\end{appendices}





\ifCLASSOPTIONcaptionsoff
  \newpage
\fi